\documentclass[journal,a4paper,10pt]{IEEEtran}
\usepackage{color}
\usepackage{cite}
\usepackage{amssymb}
\usepackage{amsmath}
\usepackage{graphicx}                  
\usepackage{bm}
\usepackage{epsfig,psfrag}
\usepackage{pst-node}
\usepackage{bardiag}

\newcommand{\be}{\begin{equation}}
\newcommand{\ee}{\end{equation}}
\newcommand{\ist}{\hspace*{.3mm}}
\newcommand{\rmv}{\hspace*{-.3mm}}
\newcommand{\iist}{\hspace*{1mm}}
\newcommand{\rrmv}{\hspace*{-1mm}}
\newcommand{\bd}[1]{\mathbf{#1}}
\newcommand{\cl}[1]{\mathcal{#1}}
\newcommand{\nn}{\nonumber}
\newcommand{\T}{\text{T}}
\newcommand{\sist}{\hspace*{.15mm}}

\allowdisplaybreaks

\begin{document}

\title{A Scalable Algorithm for Tracking an Unknown Number of Targets Using Multiple Sensors
\thanks{F.\ Meyer and P.\ Braca are with the NATO STO Centre for Maritime Research and Experimentation, 
La Spezia 19126, Italy (e-mail: \{florian.meyer, paolo.braca\}@cmre.nato.int). 
P.\ Willett is with the Department of ECE, University of Connecticut, Storrs, CT 06269, USA (e-mail: willett@engr.uconn.edu).
F.\ Hlawatsch is with the Institute of Telecommunications, TU Wien, 1040 Vienna, Austria (e-mail: franz.hlawatsch@tuwien.ac.at).
This work was supported by the 
NATO Supreme Allied Command Transformation under projects SAC000601 and SAC000608,
by the Naval Postgraduate School via ONR contract N00244-16-1-0017,
by the Austrian Science Fund (FWF) under project P27370-N30, and 
by the National Sustainability Program of the European Commission under project LO1401.
Parts of this paper were previously presented at Fusion 2015, Washington D.C., USA, July 2015
and at Fusion 2016, Heidelberg, Germany, July 2016.
}\vspace{0mm}}

\author{Florian Meyer, \emph{Member, IEEE}, 
Paolo Braca, \emph{Member, IEEE}, 
Peter Willett, \emph{Fellow, IEEE},\\ 
and Franz Hlawatsch, \emph{Fellow, IEEE}
\vspace*{-2mm}
 }

\maketitle

\begin{abstract}
We propose a method for tracking an unknown number of targets based on measurements provided by multiple sensors.
Our method achieves low computational complexity and excellent scalability 
%% in the number of targets, number of sensors, and number of measurements per sensor 
by running belief propagation 
% (BP) message passing scheme
on a suitably devised factor graph. 
A redundant formulation of data association uncertainty and the use of ``augmented target states'' including binary target indicators 
%% variables 
make it possible to exploit statistical independencies for a drastic reduction of complexity. An increase in the number of targets, sensors, or measurements leads to additional variable nodes in the factor graph but not to higher dimensions of the messages. As a consequence, the 
%% computational 
complexity of our
%% the proposed 
method scales only quadratically in the number of targets, linearly in the number of sensors,
%%  (assuming a fixed number of message passing iterations), 
and linearly in the number of measurements per sensors. The performance of the method 
%% in terms of mean optimal subpattern assignment (OSPA) error 
compares well with that of previously proposed
%% state-of-the-art 
methods, including methods with a 
less favorable scaling behavior. In particular, 
%% we observed that 
our method can outperform multisensor versions of the probability hypothesis density (PHD) filter, 
the cardinalized PHD 
%% (CPHD) 
filter, and the multi-Bernoulli 
%% (MB) 
filter. 
\vspace{.7mm}
\end{abstract}


\vspace{.7mm}

%%%%%%%%%%%%%%%%%%%%%%%%%%%%%%%%%%%%%%%%%%%%%%%%%%%%%%%%%%%
\section{Introduction}
\label{sec:intro}
%%%%%%%%%%%%%%%%%%%%%%%%%%%%%%%%%%%%%%%%%%%%%%%%%%%%%%%%%%%

\vspace{.5mm}

\subsection{Multitarget Tracking Using Multiple Sensors}
%% Background and State of the Art}
\label{sec:muttrack}
%%%%%%%%%%%%%%%%%%%%%%%%%%%%%%%%%%%%%%%%%%%%%%%%%%%%%%%%%%%

\vspace{.5mm}

Multitarget tracking is 
%% an 
important 
%% task 
in many applications including surveillance, autonomous driving, biomedical analytics, robotics, and oceanography
%% surveillance, remote sensing, air traffic control, biomedical analytics, oceanography, and computer vision 
\cite{koch14,wiley15,barShalom95,mahler2007statistical}.
%% and a key component of numerous innovative products and services \cite{koch14}.
Multitarget tracking aims at estimating the states---i.e., positions
%% , velocities, 
and possibly further parameters---of moving objects (targets) over time, 
based on measurements provided by 
%% remote 
sensing devices such as radar, sonar, or cameras \cite{barShalom95}.
Often information from multiple sensors is required to obtain 
%% a 
satisfactory 
%% level of 
reliability and 
%% tracking 
accuracy. 
%% Another complicating factor is the fact that the 
The number of targets is usually unknown \cite{mahler2007statistical} and there is a data association uncertainty, 
i.e., an unknown association between measurements and targets \cite{barShalom95}.

%% \newpage %%%%%%
 
Traditional methods for multitarget tracking model the target states as a random vector, i.e., an ordered list of random variables, 
and estimate them jointly with the random association variables. 
Examples are the joint probabilistic data association (JPDA) filter \cite{barShalom95} and the multiple hypothesis density tracker (MHT) \cite{reid79}
and their extensions to multiple sensors \cite{pao95,deb97,vermaak05}. Most of these methods assume that the number of targets is fixed and known,
%%  and the targets are ordered, which are 
which is typically not true in practice. 
Because of this assumption, most traditional methods do not solve the track management problem, i.e., they are unable to create or cancel a 
%% new 
track when a 
%% new 
target appears or disappears, respectively. 
Track management extensions of the single-sensor JPDA filter and single-sensor MHT include the joint integrated probabilistic data association (JIPDA) filter 
\cite{musicki04}, the joint integrated track splitting (JITS) filter \cite{musicki09}, and the search-initialize-track filter \cite{horridge09}.

%% \newpage %%%%%%

A more recent
%% , very promising 
class of multitarget tracking methods is based on finite set statistics (FISST). These methods calculate an approximation of the posterior multiobject probability density function (pdf), which is a joint distribution of the \emph{unordered} target states. Typically, this quantity is then used to estimate a (possibly unordered) set of target states, which is described as a random finite set.
%%  (RFS).  
%% Multitarget tracking algorithms based on FISST 
Notable examples include the probability hypothesis density (PHD) filter \cite{mahler03, mahler2007statistical, vo05}, the cardinalized PHD (CPHD) filter \cite{mahler07, mahler2007statistical, vo07}, the Bernoulli filter \cite{ristic13}, and the multi-Bernoulli (MB) filter \cite{mahler2007statistical, vo09}. These FISST-based tracking methods avoid the data association problem and implicitly perform track management. However, most existing 
%% FISST-based multitarget tracking 
methods are restricted to a single sensor. 

Even more recently, FISST-based multitarget tracking methods using labeled random finite sets have been proposed. These filters track an unknown number of targets that are identified by an (unobserved) label, and thus are able to estimate individual target tracks. In particular, the labeled multi-Bernoulli (LMB) filter \cite{reuter14} and the generalized LMB filter \cite{vo14} achieve good estimation accuracy with a computational complexity that is similar to that of the CPHD filter. Alternatively, the track-oriented marginal Bernoulli/Poisson (TOMB/P) filter and the measurement-oriented marginal Bernoulli/Poisson (MOMB/P) filter proposed in \cite{williams2015marg} can estimate individual target tracks by integrating probabilistic data association into FISST-based sequential estimation; they are not based on labeled random finite sets. 
%% However, the methods in \cite{vo14,reuter14,williams2015marg} have been formulated only for a single sensor.

In the case of low-observable targets, i.e., 
%% corresponding 
targets leading to measurements with a low signal-to-noise ratio, reliable detection and tracking using a single sensor may be impossible. 
Theoretical results \cite{braca13} suggest that the probability of detection  can be strongly improved by increasing the number of sensors. 
Unfortunately, the computational complexity of optimum multisensor-multitarget tracking 
%% methods 
scales exponentially in the number of sensors, number of targets, and number of measurements per sensor \cite{mahler09a,nannuru15,nannuru15journal,delande11}. 
Computationally feasible multisensor-multitarget tracking methods
%% \linebreak %%%%%%
include the iterator-corrector (C)PHD or briefly IC-(C)PHD filter \cite{nagappa11}, 
%% which computes measurements from different sensors sequentially, 
the approximate product multisensor (C)PHD filter \cite{mahler10a}, and the partition-based multisensor (C)PHD (MS-(C)PHD) filter \cite{nannuru15,nannuru15journal}. 
%% However, these 
These methods either use
%% involve
%% perform 
approximations of unknown fidelity and thus may not be able to fully realize the performance gains promised
%% offered 
by multiple sensors, or they still scale poorly in relevant system parameters. 
A further disadvantage of the IC-(C)PHD filter is the strong dependence of its performance on the order in which the sensor measurements are processed 
\cite{mahler10a, nagappa11, nannuru15,nannuru15journal}.  
We note that the methods in \cite{vo14,reuter14,williams2015marg} have been formulated only for a single sensor.

%% \vspace{-.5mm}

\subsection{The Proposed Method and Other Message Passing Methods}
%% Contributions and Paper Organization}
\label{sec:proposed}
%%%%%%%%%%%%%%%%%%%%%%%%%%%%%%%%%%%%%%%%%%%%%%%%%%%%%%%%%%%

\vspace{.5mm}

Here, we propose a multisensor method for multi\-target tracking with excellent scalability in the number of targets, number of sensors, 
and number of measurements per sensor. Our method allows for an unknown, time-varying number of targets (up to a specified
%% although it requires the specification of a 
\emph{maximally possible} number of targets), i.e., it implicitly performs track management. These advantages are obtained by performing ordered estimation 
using belief propagation (BP) message passing, based on the 
%% so-called 
sum-product algorithm \cite{kschischang01,wymeersch07,wainwright08,loeliger}. 
Contrary to most FISST-based methods, which calculate an approximation of the \emph{joint} posterior multiobject pdf, BP provides accurate approximations of the \emph{marginal} posterior pdfs for the individual targets. These are then used to perform Bayesian detection and estimation of the target states. 

The proposed BP method is derived by formulating a detection-estimation problem involving all the target states, 
%% target 
existence variables, and association variables---which are modeled via random vectors---for all times, targets, and sensors. We use a redundant formulation of 
data association uncertainty in terms of both target-oriented and measurement-oriented association vectors 
%% as proposed in 
\cite{chertkov10,williams14}, and
%% . Furthermore, it uses 
%% a fixed number of 
``augmented target states'' 
%% Augmented target states are hybrid states for (existing or nonexisting) targets 
%% in which the continuous state variables of existing and nonexisting targets are complemented by 
that include binary target existence indicators. 
%% To obtain a consistent Bayesian estimation problem involving the augmented target states,
%% we introduce a ``dummy pdf'' for the spatial distribution of nonexisting targets. 
%If a target does not exist, its spatial state is irrelevant and the corresponding pdf 
%% of its spatial state 
%is represented by a ``dummy pdf'' for mathematical consistency. 
%% However, 
In contrast to FISST-based techniques, the joint augmented target state is ordered and has a fixed number of components.
%% length. 

By this new formulation of the multisensor-multi\-target detection-estimation problem,
the statistical structure of the problem can be described by a factor graph, and the problem can be solved 
%% in an efficient manner 
using loopy BP message passing. The advantage of the BP approach is that it exploits conditional statistical independencies
%% \linebreak %%%%%%%% 
%% of variables 
for a drastic reduction of 
%% computational 
complexity \cite{kschischang01,wymeersch07,wainwright08,loeliger}. 
We use a ``detailed'' factor graph in which each target state and each association variable is modeled as an individual node. 
Because this factor graph involves only low-dimensional variables, the resulting BP algorithm does not perform high-dimensional operations. 
As a consequence, the complexity of our method scales only  quadratically in the number of targets, linearly in the number of sensors, 
and linearly in the number of measurements per sensors (assuming a fixed number of message passing iterations).
In addition, because our method uses
%% employs 
%% particle representations and 
particle-based calculations of all messages and beliefs, it is suited to general nonlinear, non-Gaussian measurement and state evolution models.
%% , which results in excellent scalability.

Simulation results in a challenging scenario with intersecting targets demonstrate that our method exhibits excellent scalability and, at the same time, 
its performance compares well with that of previously proposed
%% state-of-the-art 
methods. This 
%% These state-of-the-art methods include 
includes methods with a less favorable scaling behavior, namely, 
%% a scaling behavior that is 
cubic in the number of measurements and in the number of targets. In particular, our method can outperform the IC-PHD and IC-CPHD filters \cite{nagappa11}, 
the IC-MB filter \cite{vo09}, and the MS-PHD and MS-CPHD filters \cite{nannuru15,nannuru15journal}.
Furthermore, its performance does not depend on an assumed order of processing the measurements of the different sensors.

%% \subsection{Review of Message Passing Methods for Data Association and Multitarget Tracking}
%% \label{sec:mp-review}
%%%%%%%%%%%%%%%%%%%%%%%%%%%%%%%%%%%%%%%%%%%%%%%%%%%%%%%%%%%

%% \vspace{-1mm}

%% \subsection{Related Message Passing Algorithms}
%% \label{sec:related_MP}
%%%%%%%%%%%%%%%%%%%%%%%%%%%%%%%%%%%%%%%%%%%%%%%%%%%%%%%%%%%

To the best of our knowledge, previously proposed BP methods for multisensor-multitarget tracking
%% ---including data association---
are limited to the method presented in \cite{chen09} and our previous method in \cite{meyer15scalable}. In 
%% the method proposed in 
\cite{chen09}, all target states and association variables at one time step are modeled as a joint state. This results in a tree-structured
factor graph for which BP 
%% message passing 
is exact but
%% , again, 
also in an unappealing
%% poor 
scalability in the number of targets. 
%% Besides its suitability for an unknown number of targets, 
%% Our present method differs from \cite{chen09} in that it is suitable for an unknown number of targets; furthermore, it 
In contrast, our method is based on 
%% uses a previously proposed redundant BP formulation of data association 
%% that was introduced in 
%% \cite{chertkov10,williams14}. 
%% This formulation 
%% of data association 
%% leads to a ``detailed'' 
a detailed (but loopy) factor graph that gives rise to low-dimensional
%% each target state and each association variable is modeled as an individual node and consequently 
messages and, in turn, results in the attractive scaling properties described above.
%%  variables, the resulting BP algorithm does not perform any high-dimensional operations. 
%% As a consequence, the complexity of our method scales only  quadratically in the number of targets, linearly in the number of sensors, 
%% and linearly in the number of measurements per sensors (assuming a fixed number of message passing iterations).
Furthermore, both methods \cite{chen09,meyer15scalable} assume that the number of targets is known,
whereas our present method
%%  proposed here 
is suited to an unknown number of targets.

%% Various 
BP-based methods have also been 
%% previously 
proposed for 
%% pure data association, i.e., 
the problems of data association alone or data association within a multitarget tracking scheme where the tracking itself is not done by BP.
%% message passing. 
%% without multitarget tracking or with multitarget tracking outside the message passing scheme. 
In particular, BP has been used in \cite{chertkov10} and \cite{williams14} to calculate approximate marginal association probabilities for a single sensor; in \cite{horridge06} to calculate exact marginal association probabilities for a single sensor; in \cite{chen06} to calculate approximate association probabilities for multiple sensors with overlapping regions of interest; and in \cite{williams2015marg} to calculate approximate association probabilities
%%  data association 
for a single sensor. In contrast to these methods, our method uses BP for the overall multisensor-multitarget tracking problem, of which data association is only a part.

\vspace{-.5mm}

\subsection{Paper Organization}
\label{sec:org}
%%%%%%%%%%%%%%%%%%%%%%%%%%%%%%%%%%%%%%%%%%%%%%%%%%%%%%%%%%%

This paper is organized as follows. The system model and the multisensor-multitarget tracking problem are described in Section \ref{sec:systprob}, and
a statistical formulation of the problem is presented in Section \ref{sec:statform}.
In Section \ref{sec:BpShortReview}, we briefly review the framework of factor graphs and BP message passing.
Section \ref{sec:severalsensors} develops the proposed multisensor-multitarget tracking
%% BP message passing 
method. A particle-based implementation 
%% of the proposed message passing scheme 
is presented in Section \ref{sec:particleBased}. Section \ref{sec:trackManagement} proposes a scheme for choosing the birth and survival parameters. In
Section \ref{sec:existing methods}, 
%% describes 
relations of the proposed method to existing methods are discussed. 
Finally, simulation results in a scenario with intersecting targets are reported in Section \ref{sec:simres}. We note that this paper advances over the 
preliminary account of our method provided in our conference publication \cite{meyer16scalable} by adding a particle-based implementation, a scheme for choosing birth and 
survival parameters, a detailed discussion of relations to existing methods, additional performance results, and an experimental verification of scaling properties.

%% \newpage %%%%%%%

%%%%%%%%%%%%%%%%%%%%%%%%%%%%%%%%%%%%%%%%%%%%%%%%%%%%%%%%%%%
\section{System Model and Problem Statement}
\label{sec:systprob}
%%%%%%%%%%%%%%%%%%%%%%%%%%%%%%%%%%%%%%%%%%%%%%%%%%%%%%%%%%%

\vspace{.8mm}

In this section, we describe our system model and formulate the multitarget detection-estimation problem to be solved.
%% criterion used for multisensor-multitarget tracking. 

\vspace{-1mm}

\subsection{Potential Targets and Sensor Measurements}
\label{sec:pot-targ}
%%%%%%%%%%%%%%%%%%%%%%%%%%%%%%%%%%%%%%%%%%%%%%%%%%%%%%%%%%%

\vspace{.5mm}

We consider \emph{at most} $K$ targets with time-varying states. We 
%% will 
describe this situation by introducing \emph{potential targets} (PTs) $k \rmv\in\rmv \cl{K} \rmv\triangleq\rmv \{ 1,\dots,K \}$.
%% each of which exists or not.
%% the possible existence of up to $K$ targets $k \in \{ 1,\dots,K \}$ with time-varying states.
%%  $\bd{x}^{(k)}_{n}\rmv$, $n = 1,\dots,N$. 
The existence of the PTs
%%  $k$ at time $n$ 
is modeled by binary
%% Bernoulli random 
variables $r_{n,k} \rmv\in\rmv \{0,1\}$, i.e., PT $k$ exists at time $n$ if and only if $r_{n,k} \!=\! 1$. 
We also define the vector $\bd{r}_n \rmv\triangleq \big[ r_{n,1} \cdots\ist r_{n,K}\big]^\T\!$. The 
%% kinematic 
state $\bd{x}_{n,k}$ of PT $k$ at time $n$
% \in \{ 1,\dots,N \}$ 
consists of the PT's position and possibly further parameters. It will be convenient to formally consider a PT state $\bd{x}_{n,k}$ also if $r_{n,k} \!=\rmv 0$.
%% In addition, we 
We define the \emph{augmented state} 
%% consisting of the kinematic state and the existence variable, i.e., 
%% as 
$\bd{y}_{n,k} \rmv\triangleq\rmv [\bd{x}^\T_{n,k} \; r_{n,k}]^\T\rmv$ 
%% (Note that our notation indicates the augmented target state by using a sans serif font.) 
and the \emph{joint augmented state}
%%  at time $n$ is denoted 
$\bd{y}_{n} \!\triangleq\rmv \big[\bd{y}^{\T}_{n,1} \cdots\ist \bd{y}^{\T}_{n,K}\big]^{\T}\!$. 

There are
%% We consider 
$S$ sensors $s \rmv\in\rmv \cl{S} \rmv\triangleq\rmv \{ 1,\dots,S \}$ that produce ``thresholded'' measurements resulting from a detection process 
(as performed, e.g., by a radar or sonar device).
Let $\bd{z}^{(s)}_{n,m}\ist$, $m \rmv\in\rmv \cl{M}^{(s)}_n \rmv\triangleq\rmv \big\{ 1,\dots,M^{(s)}_n \big\}$ denote the measurements produced by sensor $s$ at time $n$.
We also define the stacked measurement vectors $\bd{z}^{(s)}_n \rmv\triangleq\rmv \big[\bd{z}^{(s)\T}_{n,1} \cdots\ist \bd{z}^{(s)\T}_{n,M^{(s)}_n} \big]^{\T}\rmv$ 
and $\bd{z}_n \rmv\triangleq$\linebreak %%%%%%% 
$\big[\bd{z}^{(1)\T}_n \cdots\ist \bd{z}^{(S)\T}_n \big]^{\T}\!$, and the vector $\bd{m}_n \rmv\triangleq\rmv \big[M^{(1)}_n \cdots\ist M^{(S)}_n \big]^{\T}\!$.
%% The measurements are subject to 

Because the measurements are thresholded, the multitarget tracking problem is complicated by
%%  there is 
a data association uncertainty: it is unknown which measurement $\bd{z}^{(s)}_{n,m}$ originated from which PT $k$, and
it is also possible that a measurement $\bd{z}^{(s)}_{n,m}$ did not originate from any PT (false alarm, clutter) 
or that a PT did not generate any measurement of sensor $s$ (missed detection) \cite{barShalom95,mahler2007statistical}. We make the usual assumption 
that at any time $n$, an existing target can generate at most one measurement at sensor $s$, and a measurement at sensor $s$ can be generated by at most one existing target \cite{barShalom95,mahler2007statistical}. The 
%% possible 
%% (unknown) 
%% target-
PT-measurement associations at sensor $s$ and time $n$ can then be described by the 
%% $K$-dimensional 
\emph{target-oriented association variables}
\[ 
a^{(s)}_{n,k} \ist\triangleq \begin{cases} 
    m \rmv\in\rmv \cl{M}^{(s)}_n \rmv , & \begin{minipage}[t]{50mm}at time $n$, PT $k$ generates measure-\\[-.5mm]ment $m$ at sensor $\rmv s$\end{minipage}\\[4mm]
   0 \ist, & \begin{minipage}[t]{50mm}at time $n$, PT $k$ is not detected
   %% sensed
				      by\\[-.5mm]sensor $s$.\end{minipage}
  \end{cases}
\vspace{.3mm}
\]
We also define $\bd{a}^{(s)}_{n} \rmv\triangleq\rmv \big[a^{(s)}_{n,1} \cdots\ist a^{(s)}_{n,K} \big]^{\T}\rmv$ and 
$\bd{a}_n \rmv\triangleq\rmv \big[\bd{a}^{(1)\T}_n \rmv\cdots$\linebreak %%%%%%% \ist 
$\bd{a}^{(S)\T}_n \big]^{\T}\!$. 

Following \cite{chertkov10} and \cite{williams14}, we also use an alternative
%%  redundant 
description of the PT-measurement associations 
%% at sensor $s$, which is 
in terms of the \emph{measurement-oriented association variables}
\vspace{.5mm}
\be 
b_{n,m}^{(s)} \ist\triangleq \begin{cases} 
    k \rmv\in\rmv \cl{K} \ist , & \begin{minipage}[t]{48mm}at time $n$, measurement $m$ at
    \\[-.5mm]sensor $s$ is generated by PT $k$\end{minipage}\\[4mm]
   0 \ist, & \begin{minipage}[t]{48mm}at time $n$, measurement $m$ at\\[-.5mm]
   sensor $s$ is not generated by a PT.\end{minipage}
  \end{cases}
\label{eq:b_def}
\vspace{.5mm}
\ee
We also define $\bd{b}_{n}^{(s)} \rmv\triangleq\rmv \big[b_{n,1}^{(s)} \cdots\ist b_{n,M_n^{(s)}}^{(s)} \big]^{\T}\rmv$ 
\vspace{-.3mm}
and 
$\bd{b}_n \rmv\triangleq\rmv \big[\bd{b}^{(1)\T}_n \rmv\cdots$\linebreak %%%%%%% \ist 
$\bd{b}^{(S)\T}_n \big]^{\T}\!$. 
%% We also define $\bd{b} \triangleq \big[\bd{b}_1^{\T} \cdots\ist \bd{b}^{\T}_n \big]^{\T}\rmv$. 
The description in terms of $\bd{b}_{n}^{(s)}$ is redundant in that $\bd{b}_{n}^{(s)}$ can be derived from 
%% the target-oriented association vector 
$\bd{a}^{(s)}_{n}$
%% considered previously, 
and vice versa.

\vspace{-1mm}

\subsection{Target Detection and State Estimation}
\label{sec:prob}
%%%%%%%%%%%%%%%%%%%%%%%%%%%%%%%%%%%%%%%%%%%%%%%%%%%%%%%%%%%

\vspace{.5mm}

The problem considered
%% addressed in this paper 
is detection of the PTs $k \!\in\! \cl{K}$ (i.e., of the 
%% corresponding 
binary target existence variables $r_{n,k}$) 
and 
%% Bayesian 
estimation of the target states $\bd{x}_{n,k}$ 
%% of all targets $k \in \{ 1,\dots,K \}$ 
from the past and present measurements of all the sensors $s \!\in\! \cl{S}$, i.e., 
\pagebreak %%%%%%%
from the total measurement vector 
$\bd{z} \triangleq \big[\bd{z}_1^{\T} \cdots\ist \bd{z}^{\T}_n \big]^{\T}\!$.
In a Bayesian setting, this essentially amounts to calculating the marginal posterior existence probabilities $p(r_{n,k} \!=\! 1|\bd{z})$ and the 
marginal posterior pdfs $f(\bd{x}_{n,k} | r_{n,k} \!=\! 1, \bd{z} )$. Target detection is performed by comparing $p(r_{n,k} \!=\! 1|\bd{z})$ 
to a 
%% detection 
threshold $P_{\text{th}}$, i.e., PT $k$ is considered to exist if $p(r_{n,k} \!=\! 1|\bd{z}) > P_{\text{th}}$ \cite[Ch.~2]{poor94}. 
For the detected targets $k$, an estimate of 
%% the target state 
$\bd{x}_{n,k}$ is then produced by
%% obtained by employing 
the minimum mean-square error (MMSE) estimator \cite[Ch.~4]{poor94}
\vspace{-1mm}
\begin{equation}
\hat{\bd{x}}^\text{MMSE}_{n,k} \,\triangleq \int \rmv \bd{x}_{n,k} \ist f(\bd{x}_{n,k} | r_{n,k} \!=\! 1, \bd{z} ) \ist \mathrm{d}\bd{x}_{n,k} \,.
\label{eq:mmse}
\end{equation}
This Bayesian two-stage detection-estimation procedure 
%% \cite{poor94}
%% , in which we first decide if the target exists and, in the positive case, then estimate its state, 
is also employed by the JITS method \cite{musicki09}, the JIPDA filter \cite{musicki04}, and certain FISST-based algorithms, e.g., \cite{mahler2007statistical, williams2015marg}. 
The main problem to be solved now is to find a computationally feasible recursive (sequential) calculation of $p(r_{n,k} \!=\! 1|\bd{z})$ and $f(\bd{x}_{n,k} | r_{n,k} \!=\! 1, \bd{z} )$.

%%%%%%%%%%%%%%%%%%%%%%%%%%%%%%%%%%%%%%%%%%%%%%%%%%%%%%%%%%%
\section{Statistical Formulation}
\label{sec:statform}
%%%%%%%%%%%%%%%%%%%%%%%%%%%%%%%%%%%%%%%%%%%%%%%%%%%%%%%%%%%

\vspace{.8mm}

Next, we present a statistical formulation of the system model and the multitarget detection-estimation problem.
%% criterion used for multisensor-multitarget tracking. 

\vspace{-1mm}

\subsection{Target States}
\label{sec:pot-targ_statist}
%%%%%%%%%%%%%%%%%%%%%%%%%%%%%%%%%%%%%%%%%%%%%%%%%%%%%%%%%%%

\vspace{.5mm}

While in our model a PT state $\bd{x}_{n,k}$ is formally defined also if $r_{n,k} \!=\rmv 0$, the states 
%% $\bd{x}_{n,k}$ 
of nonexisting PTs are obviously irrelevant. Accordingly, all pdfs and BP messages defined for an augmented state, $\phi(\bd{y}_{n,k}) = \phi(\bd{x}_{n,k}, r_{n,k})$, 
have the property that for $r_{n,k} \!=\rmv 0$, 
\begin{equation}
\label{eq:dummy}
\phi(\bd{x}_{n,k}, 0) \ist=\ist \phi_{n,k} \ist f_{\text{D}}(\bd{x}_{n,k}) \ist,
\end{equation}
where $f_{\text{D}}(\bd{x}_{n,k})$ is a ``dummy pdf.'' 
%% that integrates to one. 
The form \eqref{eq:dummy} must be consistent with a message multiplication operation (such as Equation \eqref{eq:messageMultiExample} in Section \ref{sec:BpShortReview}), 
in the sense that the resulting message product can still be expressed as in \eqref{eq:dummy}. This implies that the dummy pdf 
$f_{\text{D}}(\bd{x}_{n,k})$ satisfies $f_{\text{D}}^2(\bd{x}_{n,k}) = f_{\text{D}}(\bd{x}_{n,k})$ for all values of $\bd{x}_{n,k}$. Because $f_{\text{D}}(\bd{x}_{n,k})$ must also integrate to one,
it follows that $f_{\text{D}}(\bd{x}_{n,k})$ is $1$ on an arbitrary support of area/volume $1$ and $0$ outside that support.
Let 
\be
\phi(r_{n,k}) \ist\triangleq\rmv \int \rmv\phi(\bd{x}_{n,k}, r_{n,k}) \ist \mathrm{d} \bd{x}_{n,k} \ist.
\label{eq:phi-marginal}
\ee 
We then have for $r_{n,k} \!=\rmv 0$
\be
\phi(0) \ist=\rmv \int \rmv\phi(\bd{x}_{n,k}, 0) \ist \mathrm{d} \bd{x}_{n,k} \ist=\ist \phi_{n,k} \rmv \int \rmv f_{\text{D}}(\bd{x}_{n,k}) \ist \mathrm{d} \bd{x}_{n,k}
\ist=\ist \phi_{n,k} \ist.
\label{eq:phi-0}
\ee
Furthermore, using \eqref{eq:phi-marginal} and \eqref{eq:phi-0} yields
\begin{align*}
&\sum_{r_{n,k} \in \{0,1\}} \int \rmv\phi(\bd{x}_{n,k}, r_{n,k}) \ist \mathrm{d} \bd{x}_{n,k} \\[-1mm] 
&\hspace{12mm}=\rmv \int \rmv\phi(\bd{x}_{n,k}, 0) \ist \mathrm{d} \bd{x}_{n,k} + \int \rmv\phi(\bd{x}_{n,k}, 1) \ist \mathrm{d} \bd{x}_{n,k}\\[.5mm]
&\hspace{12mm}= \phi_{n,k} + \phi(1) \ist.
\end{align*}
Hence, if $\sum_{r_{n,k} \in \{0,1\}} \int \phi(\bd{x}_{n,k}, r_{n,k}) \ist \mathrm{d} \bd{x}_{n,k} = 1$, i.e., if
$\phi(\bd{x}_{n,k}, r_{n,k})$ is a true pdf in the sense of being normalized,
then $\phi_{n,k} + \phi(1) = 1$. In that case, 
\pagebreak %%%%%%%
$\phi_{n,k} = \phi(0)$ can be interpreted as a probability of nonexistence of PT $k$, i.e., of the event $r_{n,k} \rmv=\rmv 0$,
and $\phi(1)$ can be interpreted as a probability of existence of PT $k$, i.e., of the event $r_{n,k} \rmv=\! 1$.
%% and existence probabilities of PT $k$, respectively, i.e., $\phi_{n,k} = \phi(0) = \mathrm{Pr}[r_{n,k} \rmv=\rmv 0]$ and $\phi(1) = \mathrm{Pr}[r_{n,k} \rmv=\! 1]$.
%% the probability that potential target $k$, respectively, does not exist and exists at time $n$. That is,
%% \[
%% \text{Pr}[r_{n,k} = r] \ist=\ist \begin{cases}
%% \phi_{n,k} = \phi(0) \ist, & r \!=\! 0 \\[0mm]
%% \phi(1) \ist, & r \!=\! 1 \ist.
%% \end{cases}
%% \vspace{-1mm}
%% \]

%% \vspace{-1mm}

%% \subsection{Target Dynamics}
%% \label{sec:targ}
%%%%%%%%%%%%%%%%%%%%%%%%%%%%%%%%%%%%%%%%%%%%%%%%%%%%%%%%%%%

%% \vspace{.5mm}

The augmented target states $\bd{y}_{n,k} \rmv=\rmv [\bd{x}^\T_{n,k} \; r_{n,k}]^\T$ are assumed to evolve independently 
according to Markovian dynamic models \cite{barShalom95, mahler2007statistical}, and at time $n \rmv=\rmv 0$, 
they are 
%% augmented target states $\bd{y}_{0,k}$ 
assumed statistically independent across $k$
with prior pdfs $f(\bd{y}_{0,k}) \rmv=$\linebreak %%%%%%%
$f( \bd{x}_{0,k}, r_{0,k} )$. Thus, the pdf of $\bd{y}\triangleq \big[\bd{y}^{\T}_{0} \cdots\ist \bd{y}^{\T}_{n}\big]^{\T}\rmv$ factorizes 
\vspace{-.5mm}
as
\begin{align}
f( \bd{y})  
%% &= f( \bd{y}_{0}) \prod^{n}_{n'=1} f( \bd{y}_{n'} | \bd{y}_{n'-1}) \,, \label{eq:factorization_trans1}\\
&\ist= \prod^{K}_{k=1} \rmv f( \bd{y}_{0,k}) \rmv\prod^{n}_{n'=1} \!\rmv f( \bd{y}_{n'\!,k} | \bd{y}_{n'\rmv-1,k}) \ist.
\label{eq:factorization_trans2}
\end{align}
Here, the single-target augmented state transition pdf $f( \bd{y}_{n,k} | \bd{y}_{n-1,k}) = f( \bd{x}_{n,k}, r_{n,k} | \bd{x}_{n-1,k}, r_{n-1,k})$ is given as follows. 
If PT $k$ did not exist at time $n-1$, i.e., $r_{n-1,k} \!=\rmv 0$, then the probability that it exists at time $n$, i.e., $r_{n,k} \!=\! 1$, 
is given by the birth probability $p^{\sist\text{b}}_{n,k}$,
and if it does exist at time $n$, its state $\bd{x}_{n,k}$ is distributed according to the birth pdf $f_{\sist\text{b}}( \bd{x}_{n,k})$.
Thus, for $r_{n-1,k} \!=\rmv 0$, we have
\begin{align}
 f( \bd{x}_{n,k}, r_{n,k} | \bd{x}_{n-1,k}, 0) &= \begin{cases}
   ( 1 \!-\rmv p^{\sist\text{b}}_{n,k} ) \ist f_{\text{D}}( \bd{x}_{n,k}) \ist, & \!\!r_{n,k} = 0 \\[.8mm]
    p^{\sist\text{b}}_{n,k} \ist f_{\sist\text{b}}( \bd{x}_{n,k}) \ist, & \!\!r_{n,k} = 1 . 
   \end{cases} \nn\\[-.5mm]
   \label{eq:singleTargetStateTrans_0}\\[-7.5mm]
\nn
\end{align}
If PT $k$ existed at time $n-1$, i.e., $r_{n-1,k} \!=\! 1$, then the probability that it still exists at time $n$, i.e., $r_{n,k} \!=\! 1$, 
is given by the survival probability $p^{\sist\text{s}}_{n,k}$, and if it still exists at time $n$, its state $\bd{x}_{n,k}$ is distributed according to the 
state transition pdf $f( \bd{x}_{n,k} | \bd{x}_{n-1,k})$. Thus, for $r_{n-1,k} \!=\! 1$, we have
\begin{align}
f( \bd{x}_{n,k}, r_{n,k} | \bd{x}_{n-1,k}, 1) &=  
   \begin{cases}
		( 1 \!-\rmv p^{\sist\text{s}}_{n,k} ) \ist f_{\text{D}}( \bd{x}_{n,k}) \ist, & \!\!r_{n,k} = 0 \\[.5mm] 
		p^{\sist\text{s}}_{n,k} \ist f( \bd{x}_{n,k} | \bd{x}_{n-1,k}) \ist, & \!\!r_{n,k} = 1.
   \end{cases} \nn\\[-.5mm]
   \label{eq:singleTargetStateTrans_1}\\[-7.8mm]
\nn
\end{align}
A possible strategy for choosing $p^{\sist\text{b}}_{n,k}$, $p^{\sist\text{s}}_{n,k}$, and $f_{\sist\text{b}}( \bd{x}_{n,k})$ is presented in Section \ref{sec:trackManagement}.
We note that our previous work in \cite{meyer15scalable}, which assumed that the number of targets is known, 
is a special case of this setup
%% : it is reobtained by using 
that uses survival probabilities $p^{\sist\text{s}}_{n,k} \!=\! 1$ (existing targets always survive), 
birth probabilities $p^{\sist\text{b}}_{n,k} \!= 0$ (no targets are born), and initial prior pdfs $f( \bd{x}_{0,k}, r_{0,k} )$ with 
$\int f( \bd{x}_{0,k}, 
%% r_{0,k} = 
1 ) \ist \text{d}\bd{x}_{0,k} = 1$ (at time $n \rmv=\rmv 0$, all targets exist with probability $1$).
%% for all targets $k$ and all times $n$ 

%% \newpage %%%%%%

\vspace{-1.7mm}

\subsection{Sensor Measurements}
\label{sec:meas}
%%%%%%%%%%%%%%%%%%%%%%%%%%%%%%%%%%%%%%%%%%%%%%%%%%%%%%%%%%%

\vspace{.5mm}

An existing target $k$ 
%% with state $\bd{x}_{n,k}$ 
is detected by sensor $s$ (in the sense that the target generates a measurement $\bd{z}^{(s)}_{n,m}$ at sensor $s$)
with probability $P^{(s)}_{\text{d}}(\bd{x}_{n,k})$, which may depend on the target state $\bd{x}_{n,k}$. 
The number of false alarms at sensor $s$ is modeled by a Poisson probability mass function (pmf) with mean $\mu^{(s)}\rmv$, and the distribution of 
each false alarm measurement at sensor $s$ is described by the pdf $f_{\text{FA}}\big( \bd{z}^{(s)}_{n,m} \big)$ \cite{barShalom95,mahler2007statistical}.

The dependence of the measurement vector 
$\bd{z} = \big[\bd{z}_1^{\T} \cdots$\linebreak %%%%%%%
$\bd{z}^{\T}_n \big]^{\T}\rmv$ on the numbers-of-measurements vector 
$\bd{m} \triangleq \big[\bd{m}_1^{\T} \cdots$\linebreak %%%%%%%
$\bd{m}^{\T}_n \big]^{\T}\!$, 
the augmented state vector $\bd{y} \rmv=\rmv \big[\bd{y}^{\T}_{0} \cdots\ist \bd{y}^{\T}_{n}\big]^{\T}\!$, and the association vector 
$\bd{a} \triangleq \big[\bd{a}_1^{\T} \cdots \ist \bd{a}^{\T}_n \big]^{\T}\rmv$ is
described by the global likelihood function $f( \bd{z} | \bd{y}, \bd{a},\bd{m})$. With the 
\pagebreak %%%%%%%
commonly used assumption \cite{barShalom95, mahler2007statistical} 
that given $\bd{y}$, $\bd{a}$, and $\bd{m}$, the measurements 
%% $\bd{z}_{n}$ at the individual time steps $n$ and measurements 
$\bd{z}_{n}^{(s)}$ 
%% at the individual sensors $s$ 
are conditionally independent across time $n$ and sensor index $s$, 
%% the \emph{global likelihood function} $f( \bd{z} | \bd{y}, \bd{a}, \bd{m})$ factorizes as
the global likelihood function factorizes as
\be
f( \bd{z} | \bd{y}, \bd{a}, \bd{m}) 
%% &=\ist \prod^{N}_{n = 1} f\big( \bd{z}_{n} \big| \bd{y}_{n}, \bd{a}_{n} \rmv, M_{n}\big) \nn\\
\ist=\rmv \prod^{n}_{n' = 1} \prod^{S}_{s = 1} f\big( \bd{z}^{(s)}_{n'} \big| \bd{y}_{n'}, \bd{a}^{(s)}_{n'} \rmv, M^{(s)}_{n'}\big) \ist.
\label{eq:globalLikelihood}
\vspace{-1mm}
\ee
Assuming in addition that the different measurements $\bd{z}^{(s)}_{n,m}$ at 
%% any fixed 
sensor $s$ are conditionally independent given 
$\bd{y}_{n}$, $\bd{a}^{(s)}_{n}\!$, and $M^{(s)}_{n}\!$, we have the further factorization \cite{barShalom95, mahler2007statistical}
%% \vspace{-1mm}
\begin{align}
%% f( \bd{z}_{n} | \bd{y}_{n}, \bd{a}_{n}, \bd{m}_{n}) 
%% &\ist=\ist \prod^{S}_{s = 1} f\big( \bd{z}^{(s)}_{n} | \bd{y}_{n}, \bd{a}^{(s)}_{n} \rmv, \bd{m}^{(s)}_{n}\big) \label{eq:globalLikelihood}\\
%% &\ist=\ist \prod^{S}_{s = 1} \rmv
\hspace{-2mm}f\big( \bd{z}^{(s)}_{n} \big| \bd{y}_{n}, \bd{a}^{(s)}_{n} \rmv, M^{(s)}_{n}\big) &\ist=\ist \Bigg( \prod^{M^{(s)}_n}_{m = 1} \rmv\rmv f_{\text{FA}}\big( \bd{z}_{n,m}^{(s)} \big) \rmv\Bigg) \nn \\[0mm]
  &\hspace{7mm} \times\!\! \prod_{k\in \cl{D}^{(s)}_{\bd{a}_{n}, \bd{r}_{n}}}\!\!\!\rmv \frac{f\Big( \bd{z}_{n,a_{n,k}^{(s)}}^{(s)} \rmv\Big|\ist \bd{x}_{n,k} \Big)}{f_{\text{FA}}\Big( \bd{z}_{n,a_{n,k}^{(s)}}^{(s)} \Big)} \,.
\label{eq:globalLikelihood_1}\\[-6mm]
\nn
\end{align}
Here, $\cl{D}^{(s)}_{\bd{a}_{n}, \bd{r}_{n}}\!$ denotes the set of existing targets detected at sensor $s$ and time $n$, i.e., 
$\cl{D}^{(s)}_{\bd{a}_{n}, \bd{r}_{n}} \triangleq \big\{k \rmv\in\rmv \cl{K} : r_{n,k} \!=\! 1, a^{(s)}_{n,k} \!\neq 0 \big\}$.
If $\bd{z}^{(s)}_{n}$ is observed and thus fixed, $M^{(s)}_{n}$ is fixed as well and \eqref{eq:globalLikelihood_1}
%% the expression above 
can be written as
\vspace{0mm}
\begin{align}
%% f( \bd{z}_{n} | \bd{y}_{n}, \bd{a}_{n}, \bd{M}_{n}) \ist\propto\ist \prod^{S}_{s = 1} 
f\big( \bd{z}^{(s)}_{n} \big| \bd{y}_{n}, \bd{a}^{(s)}_{n} \rmv, M^{(s)}_{n}\big)
\ist=\ist C\big(\bd{z}^{(s)}_{n}\big) \prod^{K}_{k=1} g\big( \bd{x}_{n,k} , r_{n,k}, a^{(s)}_{n,k}; \bd{z}_{n}^{(s)} \big) \ist, \nn \\[-3mm]
\label{eq:factorization_like}\\[-7.5mm]
\nn
\end{align}
%% where $\propto$ denotes equality up to a normalization factor and
where $C\big(\bd{z}^{(s)}_{n}\big)$ is a normalization factor that depends only on $\bd{z}^{(s)}_{n}\!$ and $g\big( \bd{x}_{n,k} , r_{n,k}, a^{(s)}_{n,k}; \bd{z}_{n}^{(s)} \big)$ is defined 
%% \vspace{-1mm}
as
%% \vspace{-.5mm}
\begin{align}
g\big( \bd{x}_{n,k}, 1, a^{(s)}_{n,k}; \bd{z}_{n}^{(s)} \big) &= \begin{cases}
    \displaystyle\frac{f\big( \bd{z}^{(s)}_{n,m} \big| \bd{x}_{n,k} \big)}{f_{\text{FA}}\big( \bd{z}^{(s)}_{n,m} \big)} \ist, 
      & \rmv\rmv a^{(s)}_{n,k} = m \rmv\in\rmv \cl{M}^{(s)}_n\\[4mm]
     1 \ist, & \rmv\rmv a^{(s)}_{n,k} = 0 \ist
  \end{cases}\nn\\[2mm]
g\big( \bd{x}_{n,k}, 0, a^{(s)}_{n,k}; \bd{z}_{n}^{(s)} \big) &= 1.
\label{eq:likelihoodFactors} \\[-5mm]
\nn
\end{align}
%% The fact that the likelihood function in \eqref{eq:factorization_like} factorizes with respect to the target states will be important 
%% for the development of the proposed BP algorithm in Section \ref{sec:severalsensors}.
Inserting \eqref{eq:factorization_like} into \eqref{eq:globalLikelihood} yields
%%  the factorization of the \emph{global likelihood function} as
\begin{align}
f( \bd{z} | \bd{y}, \bd{a}, \bd{m}) &=\ist C(\bd{z}) \prod^{n}_{n' = 1} \prod^{S}_{s = 1} \prod^{K}_{k=1} g\big( \bd{x}_{n'\!,k} , r_{n'\!,k}, a^{(s)}_{n'\!,k}; \bd{z}_{n'}^{(s)} \big) \ist, \nn\\[-3mm]
\label{eq:globalLikelihood_3}\\[-7.5mm]
\nn
\end{align}
where $C(\bd{z})$ is a normalization factor that depends only on 
\vspace{-1mm}
$\bd{z}$.

%% \newpage %%%%%%

\subsection{Joint Prior Distribution of Association Variables and Numbers of Measurements}
\label{sec:prior}
%%%%%%%%%%%%%%%%%%%%%%%%%%%%%%%%%%%%%%%%%%%%%%%%%%%%%%%%%%%

\vspace{.5mm}

Under the assumption that given $\bd{y}$, the $\bd{a}_n^{(s)}$ and the $M_n^{(s)}$ are conditionally independent across $n$ and $s$
\cite{barShalom95, mahler2007statistical}, the joint prior pmf of the association vector $\bd{a}$ and the numbers-of-measurements vector $\bd{m}$ given 
%% the joint augmented state 
$\bd{y}$ factorizes as
\begin{equation}
p(\bd{a}, \bd{m} | \bd{y} ) \ist=\ist \prod^{n}_{n' = 1} \prod^{S}_{s = 1} 
%% \phi(\bd{a}^{(s)}_{n}) 
p\big(\bd{a}_{n'}^{(s)} \rmv, M_{n'}^{(s)} \big| \bd{y}_{n'} \big) \ist.
\label{eq:assocprior}
\vspace{-.5mm}
\end{equation}
Assuming a random
%% an unknown 
permutation of the measurements $\bd{z}^{(s)}_{n,m}$, $m \rmv\in\rmv \cl{M}^{(s)}_n$ at sensor $s$, 
with each 
%% of the $M^{(s)}_{n}!$ possible 
permutation equally likely, it is shown in \cite{barShalom95, horridge09} 
\pagebreak %%%%%%%%%
that
%% $p\big(\bd{a}^{(s)}_{n} \rmv, M^{(s)}_{n} \big| \bd{y}_{n} \big)$ factorizes as
%% For a factorization of $p\big(\bd{a}^{(s)}_{n} \rmv, M^{(s)}_{n} \big| \bd{y}_{n} \big)$, 
%% we recall that the number of false alarms at sensor $s$ is modeled by a Poisson pmf with mean $\mu^{(s)}\rmv$, and an existing target with state $\bd{x}_{n,k}$ 
%% is detected at sensor $s$ with probability $P^{(s)}_{\text{d}}(\bd{x}_{n,k})$. It has been shown in \cite{barShalom95, horridge09} that
\begin{align}
&p\big(\bd{a}^{(s)}_{n} \rmv, M^{(s)}_{n} \big| \bd{y}_{n} \big) \nn \\[1.5mm]
&\ist=\iist \psi\big(\bd{a}^{(s)}_{n}\big)\ist \frac{e^{-\mu^{(s)}} (\mu^{(s)})^{M^{(s)}_{n}-\left|\cl{D}^{(s)}_{\bd{a}_{n}, \bd{r}_{n}}\right|}}{M^{(s)}_{n}!} 
  \Bigg( \prod_{k \in \cl{D}^{(s)}_{\bd{a}_{n},\bd{r}_{n}}} \!\!\!\! P^{(s)}_{\text{d}}(\bd{x}_{n,k}) \rmv \Bigg)\nn\\[1.5mm]
&\hspace{10mm} \times \!\!\! \prod_{k'\rmv \notin \cl{D}^{(s)}_{\bd{a}_{n},\bd{r}_{n}}} \!\!\!\!\big[1\big(a^{(s)}_{n,k'}\big) - r_{n,k'} P^{(s)}_{\text{d}}(\bd{x}_{n,k'}) \big] \ist,
\label{eq:assocpriorSens}\\[-6.5mm]
\nonumber
\end{align}
where
%% \vspace{-1mm}
\be
%% \hspace*{-1.7mm}
\psi\big(\bd{a}^{(s)}_{n}\big) \triangleq \begin{cases} 
    0 \ist, &\!\! \exists \ist k,k' \!\in \cl{K}\;\ist \text{such that } a^{(s)}_{n,k} \!= a^{(s)}_{n,k'} \!\neq\rmv 0\\[.5mm]
    1 \ist, &\!\! \text{otherwise}, 
  \end{cases} 
\label{eq:mutualExclusiv} 
\ee
and $1(a)$ denotes the indicator function of the event $a \rmv=\rmv 0$ (i.e., $1(a) \rmv=\rmv 1$ if $a \rmv=\rmv 0$ and $0$ otherwise).
Note that the factor $\psi\big(\bd{a}^{(s)}_{n}\big)$ enforces
%% expresses
%% ``encodes'' 
the exclusion assumptions stated in Section \ref{sec:pot-targ}, i.e.,
that each existing target can generate at most one measurement at sensor $s$ and each measurement at sensor $s$ can be generated by at most one target. 
%% \cite{barShalom95, mahler2007statistical}. 
We
%%  the prior in \eqref{eq:assocprior} can be expressed as
can express \eqref{eq:assocpriorSens} as
%% \vspace{0mm}
\begin{align}
%% p\big(\bd{a}_{n}, \bd{m}_{n} | \bd{y}_{n} \big) 
p\big(\bd{a}^{(s)}_{n} \rmv, M^{(s)}_{n} \big| \bd{y}_{n} \big) &\ist=\ist
%% \rmv\rmv \prod^{S}_{s = 1} \rmv\rmv
C\big(M^{(s)}_{n}\big) \ist\psi\big(\bd{a}^{(s)}_{n}\big) \nn\\[.5mm]
&\hspace{7mm}\times \rmv\prod^{K}_{k=1}\rmv h\big( \bd{x}_{n,k} , r_{n,k}, a^{(s)}_{n,k}; M^{(s)}_{n} \big) \ist, \label{eq:propLikelihood} \\[-3mm]
\nn \\[-8mm]
\nn
\end{align}
where $C\big(M^{(s)}_{n}\big)$ is a normalization factor depending only on 
%% the number of measurements 
$M^{(s)}_{n}$ and $h\big( \bd{x}_{n,k} , r_{n,k}, a^{(s)}_{n,k}; M^{(s)}_{n} \big)$ is defined as
\begin{align}
h\big( \bd{x}_{n,k}, 1, a^{(s)}_{n,k}; M^{(s)}_{n} \big) &= \begin{cases}
    \displaystyle \frac{P^{(s)}_{\text{d}}(\bd{x}_{n,k})}{\mu^{(s)}} \ist, 
      & \rmv\rmv a^{(s)}_{n,k} \rmv\in\rmv \cl{M}^{(s)}_n\\[3mm]
     1 \!-\rmv P^{(s)}_{\text{d}}(\bd{x}_{n,k}) \ist, & \rmv\rmv a^{(s)}_{n,k} = 0
  \end{cases} \nn \\[2.5mm]
h\big( \bd{x}_{n,k}, 0, a^{(s)}_{n,k}; M^{(s)}_{n} \big) &= 1\big(a^{(s)}_{n,k}\big) \ist.
\label{eq:h_def_0} 
\end{align}

Using the measurement-oriented association vectors $\bd{b}_{n}^{(s)}\rmv$ defined in \eqref{eq:b_def} alongside with the target-oriented association vectors $\bd{a}^{(s)}_{n}\!$,
%% using the measurement-oriented association variables $b_{n,m}^{(s)}$, 
the exclusion-enforcing
%% indicator 
function $\psi\big(\bd{a}^{(s)}_{n}\big)$ in \eqref{eq:mutualExclusiv} can 
be formally replaced by the function 
%% $\psi\big(\bd{a}^{(s)}_{n} \rmv, \bd{b}^{(s)}_{n}\big)$ given by
%% , which factors as
%% can be written as the following product 
\cite{chertkov10, williams14} %% , vontobel13
%% \vspace{-2.5mm}
\be
\label{eq:Psi-fact}
\psi\big(\bd{a}^{(s)}_{n} \rmv, \bd{b}^{(s)}_{n}\big) = \prod_{k=1}^{K} \prod_{m=1}^{M_n^{(s)}} \!\Psi\big(a_{n,k}^{(s)} \ist,b_{n,m}^{(s)}\big) \ist,
\vspace{-1.5mm}
\ee
with
\[
%% \label{eq:Psi-fact-1}
\Psi\big(a_{n,k}^{(s)} \ist,b_{n,m}^{(s)}\big) \triangleq \begin{cases} 
    0\ist, & \begin{minipage}[t]{40mm}$a_{n,k}^{(s)} \rmv= m,\, b_{n,m}^{(s)} \rmv\neq k$\\[-.2mm]
				       \hspace*{4mm}or\;\ist$b_{n,m}^{(s)} \rmv= k,\, a_{n,k}^{(s)} \rmv\neq m$\end{minipage}\\[5.5mm]
    1\ist, & \text{otherwise}.
  \end{cases} 
\vspace{.5mm}
\]
%% Equations \eqref{eq:Psi-fact} and \eqref{eq:Psi-fact-1} taken together are equivalent to the expression \eqref{eq:mutualExclusiv} of the exclusion enforcement
%% function $\psi\big(\bd{a}^{(s)}_{n}\big)$. 
Using this redundant reformulation,
%%  of the exclusion-enforcing function, 
and defining $\bd{b} \rmv\triangleq\rmv \big[\bd{b}_1^{\T} \cdots$\linebreak 
$\bd{b}^{\T}_n \big]^{\T}\!$,
the prior pmf $p(\bd{a}, \bd{m} | \bd{y} )$ in \eqref{eq:assocprior} can be 
formally rewritten as
\vspace{-2mm}
\be
p(\bd{a}, \bd{b}, \bd{m} | \ist \bd{y} ) = \prod^{n}_{n' = 1} \prod^{S}_{s = 1} p\big(\bd{a}^{(s)}_{n'} \rmv, \bd{b}^{(s)}_{n'} \rmv, M^{(s)}_{n'} \big| \ist \bd{y}_{n'} \big) \ist,
\label{eq:factorization_prior1}
\ee
with the single-sensor prior pmfs 
\pagebreak %%%%%%%
(cf.\ \eqref{eq:propLikelihood} and \eqref{eq:Psi-fact})
\begin{align*}
&p\big(\bd{a}^{(s)}_{n} \rmv, \bd{b}^{(s)}_{n} \rmv, M^{(s)}_{n} \big| \ist \bd{y}_{n} \big) \nn \\[1mm]
&\,= C\big(M^{(s)}_{n}\big) \rmv\prod^{K}_{k=1} \rmv h\big( \bd{x}_{n,k} , r_{n,k}, a^{(s)}_{n,k}; M^{(s)}_{n}\big) 
\!\prod_{m=1}^{M_n^{(s)}} \!\rmv\Psi\big(a_{n,k}^{(s)} \ist,b_{n,m}^{(s)}\big) \ist. \nn \\[-4.5mm]
%% \label{eq:factorization_prior2}\\[-7mm]
%% \nn
\end{align*}
Thus, Equation \eqref{eq:factorization_prior1} can be expressed as
%% the factorization of the \emph{global prior} for association variables and number of measurements as
\begin{align}
p(\bd{a}, \bd{b}, \bd{m} | \ist \bd{y} ) &=\ist C(\bd{m}) \!\prod^{n}_{n' = 1} \prod^{S}_{s = 1} \prod^{K}_{k=1} \rmv h\big( \bd{x}_{n'\!,k} , r_{n'\!,k}, a^{(s)}_{n'\!,k}; M^{(s)}_{n'}\big) \nn\\[.5mm] 
&\hspace{24mm}\times \prod_{m=1}^{M_{n'}^{(s)}} \!\rmv \Psi\big(a_{n'\!,k}^{(s)} \ist,b_{n'\!,m}^{(s)}\big) \ist,
\label{eq:factorization_prior3} %% \\[-7mm]
%% \nn
\end{align}
where $C(\bd{m})$ is a normalization factor depending only on $\bd{m}$. The 
%% ``redundant'' 
factorization in \eqref{eq:factorization_prior3} constitutes
%% \pagebreak %%%%%%%
an important basis for 
%% play an important role in 
our development of the proposed 
%% efficient 
BP method in Section \ref{sec:severalsensors}.

%%%%%%%%%%%%%%%%%%%%%%%%%%%%%%%%%%%%%%%%%%%%%%%%%%%%%%%%%%%
\section{Review of 
%% Factor Graphs and\\
BP Message Passing}
\label{sec:BpShortReview}
%%%%%%%%%%%%%%%%%%%%%%%%%%%%%%%%%%%%%%%%%%%%%%%%%%%%%%%%%%%

\vspace{.5mm}

We briefly review factor graphs and the generic BP message passing scheme, which constitute the main methodological basis of the
proposed multisensor-multitarget tracking method. Consider the problem of estimating parameter 
%% (state) 
vectors $\bd{x}_k$, $k \rmv\in\rmv \{1,\ldots,K\}$ from a measurement vector $\bd{z}$. Bayesian estimation of $\bd{x}_k$ 
relies on the posterior pdf $f(\bd{x}_k|\bd{z})$ \cite{kay1993}. This pdf is a marginal pdf of the joint posterior pdf $f(\bd{x}|\bd{z})$,
where $\bd{x} = {[\bd{x}_k]}_{k=1}^K$; however, direct marginalization of $f(\bd{x}|\bd{z})$ is usually infeasible. 
An efficient marginalization can be achieved if the posterior pdf $f(\bd{x}|\bd{z})$ factorizes, i.e.,
\begin{equation}
f(\bd{x}|\bd{z}) \ist\propto\ist
%% = \frac{1}{C}
\prod_{q=1}^Q \psi_q\big( \bd{x}^{(q)}\big) \ist.
\label{eq:generalFactorization}
\end{equation}
Here, each factor argument $\bd{x}^{(q)}$ 
%% of factor $\psi_q(\cdot)$ 
comprises certain parameter vectors $\bd{x}_k$ (each 
$\bd{x}_k$ can appear in 
%% the arguments of more than one factor
several $\bd{x}^{(q)}$) and $\propto$ indicates equality up to a normalization factor. 
%% $C$ is an irrelevant normalization constant. 
Note that for compactness, our notation does not indicate the dependence of the factors $\psi_q\big( \bd{x}^{(q)}\big)$ on $\bd{z}$.

\begin{figure}[t!]
\vspace{1mm}
\centering
\psfrag{x0}[l][l][.85]{\raisebox{-3mm}{\hspace{-1.6mm}$\bd{x}_1$}}
\psfrag{x1}[l][l][.85]{\raisebox{-3.2mm}{\hspace{-1.6mm}$\bd{x}_2$}}
\psfrag{x2}[l][l][.85]{\raisebox{-3.2mm}{\hspace{-1.6mm}$\bd{x}_3$}}

\psfrag{f1}[l][l][.85]{\raisebox{-3.2mm}{\hspace{-2.2mm}$\psi_1$}}
\psfrag{f2}[l][l][.85]{\raisebox{-3.2mm}{\hspace{-1.9mm}$\psi_3$}}
\psfrag{fz1}[l][l][.85]{\raisebox{-3.8mm}{\hspace{-2.05mm}$\psi_2$}}
\includegraphics[scale=.26]{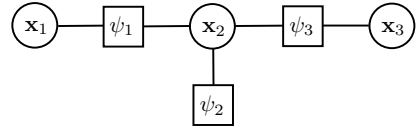}
\vspace{-1mm}
%%2.2
\caption{Factor graph representing the factorization $f(\bd{x}|\bd{z}) \propto \psi_1(\bd{x}_1,\bd{x}_2) \ist \psi_2(\bd{x}_2) \ist \psi_3(\bd{x}_2,\bd{x}_3)$.}
\label{fig:exampleFG}
\vspace{1mm}
\end{figure}

The factorization structure \eqref{eq:generalFactorization} can be represented by a \textit{factor graph} \cite{loeliger}.
As an example, for $\bd{x} = \big[\bd{x}_1^{\T} \,\ist \bd{x}_2^{\T} \,\ist \bd{x}_2^{\T} \big]^{\T}\rmv$,
the factor graph representing the factorization $f(\bd{x}|\bd{z}) \propto
%% = \frac{1}{C} \ist
\psi_1(\bd{x}_1,\bd{x}_2) \ist \psi_2(\bd{x}_2) \ist \psi_3(\bd{x}_2,\bd{x}_3)$
%% (note that $\bd{x} = \big[\bd{x}_1^{\T} \,\ist \bd{x}_2^{\T} \,\ist \bd{x}_2^{\T} \big]^{\T}$) 
is shown in Fig.\ \ref{fig:exampleFG}. In a factor graph, each parameter variable $\bd{x}_k$ is represented by a variable node and 
each factor $\psi_q(\cdot)$ 
%% is represented 
by a factor node (depicted in Fig.\ \ref{fig:exampleFG} by a circle and a square, respectively).
Variable node ``$\bd{x}_k$'' and factor node ``$\psi_q$'' are \emph{adjacent}, i.e., connected by an edge, 
if the variable $\bd{x}_k$ is an argument of the factor $\psi_q(\cdot)$.

\textit{Belief propagation (BP)}, also known as the \textit{sum-product algorithm} \cite{kschischang01}, is 
%% a message passing algorithm that is 
based on a factor graph and aims at computing the marginal posterior pdfs 
\pagebreak %%%%%%%
$f(\bd{x}_k|\bd{z})$ 
%% on a factor graph 
in an efficient way. 
%% The main principle of BP is to 
For each node, certain messages are calculated, each of which is passed to
%% exchanges these messages with 
one of the adjacent nodes.
%%  (i.e., the nodes connected via an edge).
For each variable node, the incoming and outgoing messages are functions of the corresponding variable. 
More specifically, consider a variable node ``$\bd{x}_k$'' and an adjacent factor node ``$\psi_q$'', i.e.,
%% and note that 
the variable $\bd{x}_k$ is part of the argument $\bd{x}^{(q)}$ of $\psi_q\big( \bd{x}^{(q)}\big)$. 
Then, the message passed from factor node ``$\psi_q$''
to variable node ``$\bd{x}_k$''  is given by
\vspace{-1mm}
\be
\zeta_{\psi_q \rightarrow \bd{x}_k}(\bd{x}_k) = \int \rmv\psi_q\big( \bd{x}^{(q)} \big) \!\prod_{k'\rmv\neq k} \!\eta_{\ist \bd{x}_{k'} \rightarrow \psi_q}(\bd{x}_{k'}) \, \mathrm{d}\bd{x}_{\bar{k}} \,,
\label{eq:incomingMessage}
\ee
where $\prod_{k'\rmv\neq k} \rmv \eta_{\ist \bd{x}_{k'} \rightarrow \psi_q}(\bd{x}_{k'})$ denotes the product of the messages passed to factor node ``$\psi_q$''
from all adjacent variable nodes except ``$\bd{x}_k$'', and $\int \ldots\, \mathrm{d}\bd{x}_{\bar{k}}$ denotes integration with respect to all constituent vectors of 
$\bd{x}^{(q)}$ except $\bd{x}_k$. For example, the message passed from factor node ``$\psi_1$'' to variable node $``\bd{x}_2$'' in Fig.\ \ref{fig:exampleFG} is 
$\zeta_{\psi_1 \rightarrow \bd{x}_2}(\bd{x}_2) = \int \psi_1(\bd{x}_1, \bd{x}_2) \ist \eta_{\ist \bd{x}_1 \rightarrow \psi_1}(\bd{x}_1) \ist \mathrm{d} \bd{x}_1$; note that
$\bd{x}^{(1)} = \big[\bd{x}_1^{\T} \,\ist \bd{x}_2^{\T} \big]^{\T}\!$.
The message $\eta_{\ist \bd{x}_{k} \rightarrow \psi_q}(\bd{x}_{k})$ passed from variable node ``$\bd{x}_k$'' to factor node ``$\psi_q$'' is given by the product 
of the messages passed to variable node ``$\bd{x}_k$'' from all adjacent factor nodes except ``$\psi_q$''. For example, in Fig.\ \ref{fig:exampleFG},
the message passed from variable node ``$\bd{x}_2$'' to factor node ``$\psi_1$'' is 
%% given by
$\eta_{\ist \bd{x}_2\rightarrow \psi_1}(\bd{x}_2) =  \zeta_{\psi_2 \rightarrow \bd{x}_2}(\bd{x}_2) \ist \zeta_{\psi_3 \rightarrow \bd{x}_2}(\bd{x}_2)$.
Message passing is started at variable nodes with only one edge (which pass a constant message) and/or factor nodes with only one edge 
(which pass the corresponding factor). Note that BP can also be applied to factorizations involving discrete variables by replacing integration with summation in  \eqref{eq:incomingMessage}.

Finally, for each variable node ``$\bd{x}_k$'', a \emph{belief} $\tilde{f}(\bd{x}_k)$ is calculated as the product of all incoming messages (passed from all adjacent factor nodes)
followed by a normalization such that $\int \tilde{f}(\bd{x}_k) \ist \mathrm{d}\bd{x}_k = 1$.
%%  ``$\psi_q$''
For example, in Fig.\ \ref{fig:exampleFG}, 
\begin{equation}
\label{eq:messageMultiExample}
\tilde{f}(\bd{x}_2) \propto \ist \zeta_{\psi_1 \rightarrow \bd{x}_2}(\bd{x}_2) \ist \zeta_{\psi_2 \rightarrow \bd{x}_2}(\bd{x}_2) \ist \zeta_{\psi_3 \rightarrow \bd{x}_2}(\bd{x}_2) \ist.
\end{equation}
If the factor graph is a tree, i.e., without loops, then the belief $\tilde{f}(\bd{x}_k)$ is exactly equal to the marginal posterior pdf $f(\bd{x}_k|\bd{z})$. 
For factor graphs with loops, BP is applied in an iterative manner, and the beliefs $\tilde{f}(\bd{x}_k)$ are only approximations of the respective marginal posterior pdfs $f(\bd{x}_k|\bd{z})$; these approximations have been observed to be very accurate in many applications \cite{kschischang01,wymeersch07,wainwright08}.
In this iterative ``loopy BP'' scheme, there is no canonical
%% general 
order in which the messages should be calculated, and different orders may lead to different beliefs. The choice of an appropriate order of message calculation 
will be an important aspect in our development of the proposed method.
%% multisensor-multitarget tracking algorithm. 

\vspace{-1mm}

%%%%%%%%%%%%%%%%%%%%%%%%%%%%%%%%%%%%%%%%%%%%%%%%%%%%%%%%%%%
\section{The Proposed BP-based Multisensor-\\[-.3mm] 
Multitarget Tracking Method}
%% Multi-Target Tracking Algorithm using Multiple Sensors}
\label{sec:severalsensors}
%%%%%%%%%%%%%%%%%%%%%%%%%%%%%%%%%%%%%%%%%%%%%%%%%%%%%%%%%%%

\vspace{.5mm}

%% We now develop the proposed BP-based multisensor-multitarget tracking algorithm.
The marginal posterior existence probability $p(r_{n,k} \!=\! 1|\bd{z})$ underlying target detection as discussed in Section \ref{sec:prob}
can be obtained from the marginal posterior pdf of the augmented target state, $f(\bd{y}_{n,k} | \bd{z}) = f(\bd{x}_{n,k}, r_{n,k}|\bd{z})$, according to
\begin{equation}
p(r_{n,k} \!=\! 1|\bd{z}) = \int f(\bd{x}_{n,k}, r_{n,k} \!=\! 1 |\bd{z}) \ist \text{d}\bd{x}_{n,k} \ist ,
\label{eq:margpost_r}
\end{equation}
and the marginal posterior pdf  $f(\bd{x}_{n,k} | r_{n,k} \!=\! 1, \bd{z} )$ underlying MMSE state estimation (see \eqref{eq:mmse})
can be obtained from $f(\bd{x}_{n,k}, r_{n,k}|\bd{z})$ according to
\begin{equation}
f(\bd{x}_{n,k}|r_{n,k} \!=\! 1, \bd{z} ) =  \frac{f(\bd{x}_{n,k}, r_{n,k} \!=\! 1 |\bd{z})}{p(r_{n,k} \!=\! 1|\bd{z})} \,.
\label{eq:margpost_x}
\vspace{.8mm}
\end{equation}
%% Thus, our goal is to develop an 
An efficient approximate calculation of 
%% the marginal posterior pdf 
$f(\bd{x}_{n,k}, r_{n,k}|\bd{z})$ can be obtained by performing BP message passing 
%% with an appropriate scheduling of message computations 
on a 
%% specific 
factor graph that expresses the factorization of the joint posterior pdf involving all relevant parameters.
This factor graph will be derived next.

\vspace{-1mm}

\subsection{Joint Posterior pdf and Factor Graph}
\label{sec:severalsensors_fact}
%%%%%%%%%%%%%%%%%%%%%%%%%%%%%%%%%%%%%%%%%%%%%%%%%%%%%%%%%%%

\vspace{.5mm}

%% Our goal is to compute 
The marginal posterior pdf $f(\bd{y}_{n,k} | \bd{z} ) = f(\bd{x}_{n,k}, r_{n,k}|\bd{z})$ is a marginal density of the joint posterior pdf $f(\bd{y}, \bd{a}, \bd{b} | \bd{z} )$,
%% . Here, $f(\bd{y}, \bd{a}, \bd{b} | \bd{z} )$ 
which involves all the augmented states, all the target-oriented and measurement-oriented
association variables, and all the measurements of all sensors, at all times up to the current time $n$. In the following
%% For the following 
derivation of 
%% this joint posterior pdf 
$f(\bd{y}, \bd{a}, \bd{b} | \bd{z} )$, the measurements $\bd{z}$ are 
observed and thus fixed, and consequently $M^{(s)}_{n}$ and $\bd{m}$ are fixed as well.
%% the observed measurements and the corresponding vector, respectively, i.e., they are consistent with $\bd{z}$ and fixed.
Then, using Bayes' rule 
%% in the second line 
and the fact that $\bd{a}$ implies $\bd{b}$, we obtain
\begin{align*}
f(\bd{y}, \bd{a}, \bd{b}| \bd{z} )\, &=\, f(\bd{y}, \bd{a}, \bd{b}, \bd{m}| \bd{z} )\nn\\[0mm]
&\propto\,  f( \bd{z} | \bd{y}, \bd{a}, \bd{b}, \bd{m} ) \ist f(\bd{y}, \bd{a}, \bd{b}, \bd{m})\nn\\[0mm]
&=\, f( \bd{z} | \bd{y}, \bd{a}, \bd{m} ) \, p(\bd{a}, \bd{b}, \bd{m} | \bd{y}) \ist f(\bd{y}) \ist. \nn \\[-4.5mm]
\end{align*}
%% Here, we used Bayes' rule and the fact that $\bd{a}$ implies $\bd{b}$.
%%  in the third line.
Inserting \eqref{eq:factorization_trans2} for $f(\bd{y})$, \eqref{eq:globalLikelihood_3} for $f( \bd{z} | \bd{y}, \bd{a}, \bd{m} )$, 
and \eqref{eq:factorization_prior3} for $p(\bd{a}, \bd{b}, \bd{m} | \bd{y})$ then yields 
%% and assuming that the target states are independent at time $n=0$ 
the final factorization
%% \vspace{.7mm}
\begin{align}
\hspace{-1mm}f(\bd{y}, \bd{a}, \bd{b}| \bd{z} ) & \ist\propto \prod^{K}_{k = 1} \rmv f( \bd{y}_{0,k}) \rmv\prod^{n}_{n'= 1} \! f( \bd{y}_{n'\!,k} | \bd{y}_{n'\rmv-1,k}) \nn \\[-1.7mm]
&\hspace{5.5mm} \times \prod^{S}_{s= 1} \rmv\upsilon \big( \bd{y}_{n'\!,k}, a^{(s)}_{n'\!,k}; \bd{z}^{(s)}_{n'} \big) 
  \rmv\prod^{M^{(s)}_{n'}}_{m= 1} \!\rmv \Psi\big( a^{(s)}_{n'\!,k}, b^{(s)}_{n'\!,m}  \big) \ist, \nn \\[-2.5mm]
\label{eq:factorOverall} \\[-8.5mm]
\nn
\end{align}
with 
\begin{align}
\upsilon\big( \bd{y}_{n,k}, a^{(s)}_{n,k}; \bd{z}^{(s)}_{n}  \big) 
%% &\ist\triangleq\ist 
&\triangleq\ist g\big( \bd{x}_{n,k} , r_{n,k}, a^{(s)}_{n,k}; \bd{z}_{n}^{(s)} \big) \nn\\[0mm] 
& \hspace{10mm} \times 
h\big( \bd{x}_{n,k} , r_{n,k},a^{(s)}_{n,k}; M^{(s)}_{n} \big) \ist.
%%  \ist.
\label{eq:upsilon_def}
\end{align}
This factorization 
%% \eqref{eq:factorOverall} is 
can be represented graphically by a factor graph as explained in Section \ref{sec:BpShortReview}.
This factor graph is depicted for one time step in Fig.\ \ref{fig:overallFG}; it provides the starting-point for our development of the proposed BP method. 
%% for multitarget detection and estimation in the next section.

\vspace{-1mm}

\subsection{BP Method} %% Message Passing Algorithm
\label{sec:severalsensors_bp}
%%%%%%%%%%%%%%%%%%%%%%%%%%%%%%%%%%%%%%%%%%%%%%%%%%%%%%%%%%%

\vspace{.3mm}

\begin{figure}
%% \vspace{1mm}
\centering
\psfrag{A}[l][l][1.3]{\raisebox{-3mm}{\hspace{0mm}$\bd{a}$}}
\psfrag{F1A}[l][l][1.3]{\raisebox{-2mm}{\hspace{1.5mm}$p$}}

\psfrag{D1a}[l][l][.75]{\raisebox{-4mm}{\hspace{1mm}$\beta_1$}}
\psfrag{D3a}[l][l][.75]{\raisebox{-4mm}{\hspace{0mm}$\beta_K$}}

\psfrag{E1}[l][l][.75]{\raisebox{-3mm}{\hspace{0mm}$\eta_1$}}
\psfrag{E3}[l][l][.75]{\raisebox{-2mm}{\hspace{0mm}$\eta_K$}}

\psfrag{a1}[l][l][.85]{\raisebox{-2mm}{\hspace{-.2mm}$a_1$}}
\psfrag{a3}[l][l][.85]{\raisebox{-1.8mm}{\hspace{-.7mm}$a_K$}}

\psfrag{b1}[l][l][.85]{\raisebox{-3mm}{\hspace{.2mm}$b_1$}}
\psfrag{b2}[l][l][.85]{\raisebox{-2.7mm}{\hspace{0.1mm}$b_2$}}
\psfrag{b4}[l][l][.85]{\raisebox{-2.7mm}{\hspace{-.5mm}$b_M$}}

\psfrag{F1}[l][l][.85]{\raisebox{-4mm}{\hspace{0mm}$f_1$}}
\psfrag{F3}[l][l][.85]{\raisebox{-4mm}{\hspace{-.2mm}$f_K$}}

\psfrag{B1}[l][l][.85]{\raisebox{2.2mm}{\hspace{-.2mm}$\tilde{f}_1$}}
\psfrag{B3}[l][l][.85]{\raisebox{1.5mm}{\hspace{-.5mm}$\tilde{f}_K$}}

\psfrag{na}[l][l][.85]{\raisebox{6mm}{\hspace{-.5mm}}}

\psfrag{s1}[l][l][.75]{\raisebox{-3.5mm}{\hspace{-1.4mm}$s=1$}}
\psfrag{s2}[l][l][.85]{\raisebox{0mm}{\hspace{0mm}}}
\psfrag{sK}[l][l][.75]{\raisebox{5mm}{\hspace{-1.5mm}$s=S$}}

\psfrag{P1}[l][l][.75]{\raisebox{-3mm}{\hspace{-.5mm}$\alpha_1$}}
\psfrag{P3}[l][l][.75]{\raisebox{-3mm}{\hspace{-.6mm}$\alpha_K$}}

\psfrag{FZ1a}[l][l][.75]{\raisebox{-2.5mm}{\hspace{3.6mm}$\upsilon_{1}$}}
\psfrag{FZ3a}[l][l][.75]{\raisebox{-2mm}{\hspace{2.3mm}$\upsilon_{K}$}}

\psfrag{FZ1}[l][l][.75]{\raisebox{-3mm}{\hspace{.1mm}$\upsilon_{1}$}}
\psfrag{FZ3}[l][l][.75]{\raisebox{-3mm}{\hspace{.2mm}$\upsilon_{K}$}}

\psfrag{F1a}[l][l][.85]{\raisebox{-3mm}{\hspace{-.5mm}$f_1$}}
\psfrag{F3a}[l][l][.85]{\raisebox{-3mm}{\hspace{-.5mm}$f_K$}}

\psfrag{V1a}[l][l][.85]{\raisebox{-2.5mm}{\hspace{1.2mm}$\bd{y}_1$}}
\psfrag{V3a}[l][l][.85]{\raisebox{-2mm}{\hspace{.5mm}$\bd{y}_K$}}

\psfrag{A}[l][l][.85]{\raisebox{-3mm}{\hspace{.2mm}$\bd{a}$}}
\psfrag{F1A}[l][l][.85]{\raisebox{-2mm}{\hspace{1.8mm}$p$}}

\psfrag{K1}[l][l][.75]{\raisebox{-2.5mm}{\hspace{-.5mm}$\alpha_1$}}
\psfrag{K3}[l][l][.75]{\raisebox{-2.5mm}{\hspace{-1mm}$\alpha_K$}}

\psfrag{K1a}[l][l][.75]{\raisebox{-2.5mm}{\hspace{-.3mm}$\alpha_1$}}
\psfrag{K3a}[l][l][.75]{\raisebox{-2.5mm}{\hspace{-.8mm}$\alpha_K$}}

\psfrag{S1S}[l][l][.75]{\raisebox{-2.5mm}{\hspace{1mm}$\gamma_{1}$}}
\psfrag{S3S}[l][l][.75]{\raisebox{-2mm}{\hspace{1mm}$\gamma_{K}$}}

\psfrag{S1Sa}[l][l][.75]{\raisebox{3mm}{\hspace{2mm}$\gamma_{1}$}}
\psfrag{S3Sa}[l][l][.75]{\raisebox{4.5mm}{\hspace{2mm}$\gamma_{K}$}}

\psfrag{psi1}[l][l][.75]{\raisebox{1mm}{\hspace{-4mm}$\Psi_{1,1}$}}
\psfrag{psi2}[l][l][.75]{\raisebox{-4mm}{\hspace{-1mm}$\Psi_{K,M}$}}
\psfrag{psi3}[l][l][.75]{\raisebox{-4mm}{\hspace{-2mm}$\Psi_{1,M}$}}
\psfrag{psi4}[l][l][.75]{\raisebox{-1.5mm}{\hspace{-1.8mm}$\Psi_{K,1}$}}

\psfrag{M1}[l][l][.65]{\raisebox{-3.5mm}{\hspace{-2.5mm}$\nu_{1,1}$}}
\psfrag{M1a}[l][l][.65]{\raisebox{-6mm}{\hspace{-3.5mm}$\zeta_{1,1}$}}

\psfrag{M2}[l][l][.65]{\raisebox{-3.2mm}{\hspace{-1.8mm}$\nu_{M,1}$}}
\psfrag{M2a}[l][l][.65]{\raisebox{-5mm}{\hspace{-.5mm}$\zeta_{K,1}$}}

\psfrag{M3}[l][l][.65]{\raisebox{-2.5mm}{\hspace{-1.5mm}$\nu_{1,K}$}}
\psfrag{M3a}[l][l][.65]{\raisebox{-2mm}{\hspace{0mm}$\zeta_{1,M}$}}

\psfrag{M4}[l][l][.65]{\raisebox{-2.5mm}{\hspace{-1.5mm}$\nu_{M,K}$}}
\psfrag{M4a}[l][l][.65]{\raisebox{2mm}{\hspace{-3mm}$\zeta_{K,M}$}}
\hspace{-2mm}\includegraphics[scale=.82]{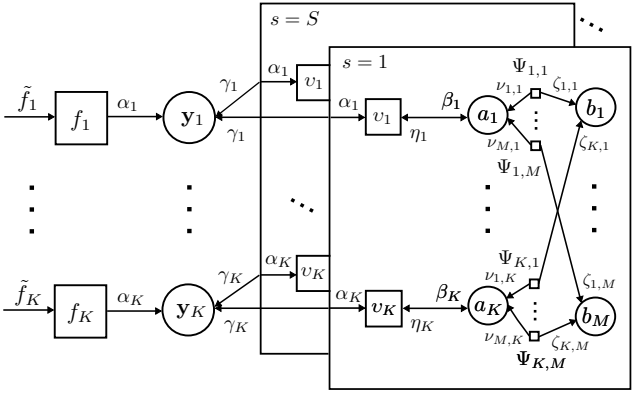}
\vspace{-1mm}
\caption{Factor graph representing the factorization of the joint posterior pdf $f(\bd{y}, \bd{a}, \bd{b}| \bd{z} )$ in \eqref{eq:factorOverall},
%%  underlying the proposed algorithm, 
depicted for one time step. For simplicity, the time index $n$ and sensor index $s$ are omitted, and the following short notations are used:
$f_{k} \rmv\triangleq f(\bd{y}_{n,k}|\bd{y}_{n-1,k})$, 
$\upsilon_{k} \rmv\triangleq\rmv \upsilon\big( \bd{y}_{n,k}, a^{(s)}_{n,k} ; \bd{z}_{n}^{(s)} \big)$, 
$\tilde{f}_k \rmv\triangleq \tilde{f}(\bd{y}_{n,k})$, 
$\Psi_{k,m} \rmv\triangleq \Psi\big(a_{n,k}^{(s)} \ist,b_{n,m}^{(s)}\big)$, 
$\alpha_{k} \rmv\triangleq \alpha (\bd{y}_{n,k})$,
$\beta_{k} \rmv\triangleq \beta \big(a^{(s)}_{n,k}\big)$, $\eta_{k} \rmv\triangleq \eta \big(a^{(s)}_{n,k}\big)$, 
$\gamma_{k} \rmv\triangleq \gamma^{(s)} (\bd{y}_{n,k})$, 
$\nu_{m,k} \rmv\triangleq  \nu^{(p)}_{m \rightarrow k}\big(a^{(s)}_{n,k}\big)$, and 
$\zeta_{k,m} \rmv\triangleq \zeta^{(p)}_{k \rightarrow m}\big(b^{(s)}_{n,m}\big)$.}
\label{fig:overallFG}
\vspace{-1mm}
\end{figure}

As discussed
%% mentioned 
in Section \ref{sec:BpShortReview}, approximations $\tilde{f}(\bd{y}_{n,k})$ of the marginal posterior pdfs 
$f(\bd{y}_{n,k} | \bd{z} )$
%% ---so-called beliefs---
%% of the augmented states $\bd{y}_{n,k}$ 
can be obtained in an efficient way by running iterative BP message passing \cite{kschischang01,wainwright08,loeliger} 
on the factor graph in Fig.\ \ref{fig:overallFG}. Since this factor graph is loopy, we have to decide on a specific order of message computation. 
We choose this order according to the following rules: (i) Messages are not sent backward 
in 
time\footnote{This %%%%%%%%
is equivalent to the approximative assumption that the target states are conditionally
%% mutually 
independent given
%% conditional on 
the past measurements, as is done in the derivation of the JPDA filter \cite{barShalom95}.} %%%%%%%%
\cite{wymeersch09}.
\pagebreak %%%%%%%%%
%% \footnote{In the tracking literature, this type of methods is known as scan-by-scan trackers \cite{barShalom95}.}
%% (ii) No iteration
%% At any given
%% a specific 
%% time step, we do not iterate
%% across different sensors is performed. (iii) 
(ii) Iterative message passing is only performed at each time step and at each sensor separately---i.e., in particular, for the loops connecting different sensors 
we only perform a single message passing iteration---and only for data association. With these rules, 
the generic BP rules for calculating messages and beliefs as summarized in Section \ref{sec:BpShortReview} yield the following BP message passing operations 
at time $n$.

First, a \emph{prediction} step is performed for all PTs $k \in \cl{K}$, 
\vspace{1mm}
i.e.,
\begin{align*}
%\label{eq:bp_predmess_s}
\alpha(\bd{x}_{n,k}, r_{n,k}) &= \rrmv \sum_{r_{n-1,k} \in \{0,1\}}\int \! f( \bd{x}_{n,k}, r_{n,k} | \bd{x}_{n-1,k}, r_{n-1,k})\nn\\[-1.5mm]
&\hspace{23.4mm} \times \tilde{f}(\bd{x}_{n-1,k}, r_{n-1,k}) \, \mathrm{d}\bd{x}_{n-1,k} \ist.\nn \\[-4.5mm] \nn
\end{align*}
Here, 
%% $\tilde{f}(\bd{y}_{n-1,k}) \triangleq 
$\tilde{f}(\bd{x}_{n-1,k}, r_{n-1,k})$ was calculated at the previous time $n\!-\!1$. 
Inserting 
%% the expressions 
\eqref{eq:singleTargetStateTrans_0} and \eqref{eq:singleTargetStateTrans_1} for $f( \bd{x}_{n,k}, r_{n,k} | \bd{x}_{n-1,k}, r_{n-1,k})$,
%%  for the single-target state transition function, 
we obtain the following expressions of $\alpha(\bd{x}_{n,k}, r_{n,k})$: 
%% the prediction step can be rewritten as follows: 
for $r_{n,k} \rmv=\rmv 1$,
\begin{align}
\hspace{-2mm}\alpha(\bd{x}_{n,k},1) &\ist=\ist p^{\sist\text{b}}_{n,k} \ist f_{\sist\text{b}}( \bd{x}_{n,k}) \ist \tilde{f}_{n-1,k} +\ist p^{\sist\text{s}}_{n,k}  \nn \\[1mm]
&\hspace{4mm} \times\! \int \! f( \bd{x}_{n,k} | \bd{x}_{n-1,k}) \tilde{f}(\bd{x}_{n-1,k},1) \, \mathrm{d}\bd{x}_{n-1,k} \ist, \label{eq:prediction_1} \\[-5mm]
\nn
\end{align}
and for $r_{n,k} \rmv=\rmv 0$,
\begin{align}
\alpha_{n,k} &\ist=\ist \big(1 \!-\rmv p^{\sist\text{b}}_{n,k}\big) \tilde{f}_{n-1,k} \ist+ \big(1\!-\rmv p^{\sist\text{s}}_{n,k}\big) \rmv\rmv\rmv\rmv  \int \rmv\rmv \!\tilde{f}(\bd{x}_{n-1,k},1) 
\,\mathrm{d}\bd{x}_{n-1,k} \ist. \nn\\[-2mm]
\label{eq:prediction_0}\\[-7mm]\nn
\end{align}
We note
%% recall 
that  
%% $\tilde{f}_{n,k} =\tilde{f}(r_{n,k}) \big|_{r_{n,k} = 0}$ and $\alpha_{n,k} = \alpha(r_{n,k}) \big|_{r_{n,k} = 0}$, where 
%% $\tilde{f}(r_{n,k}) \triangleq \int \rmv\tilde{f}(\bd{x}_{n,k}, r_{n,k}) \ist \mathrm{d} \bd{x}_{n,k}$ 
$\tilde{f}_{n-1,k} = \int \rmv\tilde{f}(\bd{x}_{n-1,k}, 0) \ist \mathrm{d} \bd{x}_{n-1,k}$ and 
%% $\alpha(r_{n,k}) \triangleq \int \rmv\alpha(\bd{x}_{n,k}, r_{n,k}) \ist \mathrm{d} \bd{x}_{n,k}$ 
$\alpha_{n,k} = \int \rmv\alpha(\bd{x}_{n,k}, 0) \ist \mathrm{d} \bd{x}_{n,k}$ (cf.\ \eqref{eq:phi-0}).
%%  and \eqref{eq:phi-marginal}). 
Furthermore, 
%% Also note that 
since $\tilde{f}(\bd{x}_{n,k},r_{n,k})$ is normalized, so is
$\alpha(\bd{x}_{n,k},r_{n,k})$, 
%% also normalizes to one, 
i.e., $\sum_{r_{n,k} \in \{0,1\}} \int \rmv\alpha(\bd{x}_{n,k},r_{n,k}) \ist \mathrm{d}\bd{x}_{n,k} \rmv=\rmv 1$. Thus, 
%% the second line in 
%% \eqref{eq:prediction} ???can also be expressed as 
we have $\alpha_{n,k} = 1 - \int \rmv\alpha(\bd{x}_{n,k},1) \ist \text{d}\bd{x}_{n,k}$. 
%% ???WHY IMPORTANT???

%% \newpage %%%%%%%

After the prediction step, the following steps are performed for all PTs $k \in \cl{K}$ and all sensors $s \in \cl{S}$ in 
parallel:

\vspace{.7mm}

\begin{enumerate}
\item \emph{Measurement evaluation}:
\begin{align}
\hspace{-2mm}\beta\big(a^{(s)}_{n,k}\big) &=\rmv\rmv \sum_{r_{n,k} \in \{0,1\}} \int \rmv\rmv \upsilon\big( \bd{x}_{n,k}, r_{n,k}, a^{(s)}_{n,k} ; \bd{z}_{n}^{(s)} \big) \nn\\[-2.5mm]
&\hspace{27mm}\times \alpha(\bd{x}_{n,k}, r_{n,k}) \, \mathrm{d}\bd{x}_{n,k} \nn\\[.5mm]
&= \int \rmv\rmv \upsilon\big( \bd{x}_{n,k}, 1, a^{(s)}_{n,k}; \bd{z}_{n}^{(s)} \big) \ist \alpha(\bd{x}_{n,k}, 1) \, \mathrm{d}\bd{x}_{n,k} \nn\\[-.3mm]
&\hspace{36mm}+\ist 1\big(a^{(s)}_{n,k}\big) \, \alpha_{n,k} \ist. \!\!
\label{eq:bp_measevalution2}
\end{align}
In the last expression, we used 
%% the fact that 
\vspace{-.5mm}
$\upsilon\big( \bd{x}_{n,k}, 0, a^{(s)}_{n,k} ; \bd{z}_{n}^{(s)} \big) = 1\big(a^{(s)}_{n,k}\big)$, which follows from 
\eqref{eq:upsilon_def} with \eqref{eq:likelihoodFactors} and
\vspace{1mm}
\eqref{eq:h_def_0}.

\item \emph{Iterative data association} (this part of the BP method
%% iterative message passing algorithm 
closely follows \cite{chertkov10, vontobel13, williams14}).
In iteration 
$p \in \{1,\dots,$\linebreak %%%%%%%%
$P\}$,
the following calculations are performed for all measurements 
\vspace{.5mm}
$m \rmv\in\rmv \cl{M}^{(s)}_n$:
\begin{align}
%% \hspace{-3mm}
&\nu^{(p)}_{m \rightarrow k}\big(a^{(s)}_{n,k}\big) \ist=\rmv \sum_{b^{(s)}_{n,m}} \!\Psi\big(a^{(s)}_{n,k}\ist,b^{(s)}_{n,m}\big) 
  \!\!\rmv\prod_{k' \in \cl{K} \backslash \{k\}}\!\!\!\rmv \zeta^{(p-1)}_{k' \rightarrow m}\big(b^{(s)}_{n,m}\big) \nn \\[-3.5mm]
&\label{eq:bp_iterativeDataAssoc_nu_1}\\[-10mm]
&\nn
\end{align}
and
%% \vspace{1mm}
\begin{align}
\hspace{-3mm}\zeta^{(p)}_{k\rightarrow m}\big(b^{(s)}_{n,m}\big) &\ist= \sum_{a^{(s)}_{n,k}} \beta\big(a^{(s)}_{n,k}\big) 
  \ist \Psi\big(a^{(s)}_{n,k}\ist,b^{(s)}_{n,m}\big) \nn\\[-2mm]
\hspace{-3mm}&\hspace{11.5mm}\times\hspace{-2mm} \prod_{m' \in \cl{M}_n^{(s)} \backslash \{m\}}\!\!\! \nu^{(p)}_{m' \rightarrow k}\big(a^{(s)}_{n,k}\big) \ist.
  \label{eq:iterativeDataAssocL}\\[-5.5mm]
\hspace{-3mm}&\nn
\end{align}
Here, $\sum_{b^{(s)}_{n,m}}$ is short for $\sum_{b^{(s)}_{n,m} \in \{0,\ldots,K\}}$ and $\sum_{a^{(s)}_{n,k}}$ is short for $\sum_{a^{(s)}_{n,k} \in \{0,\ldots,M^{(s)}_n\}}$.
The operations \eqref{eq:bp_iterativeDataAssoc_nu_1} and \eqref{eq:iterativeDataAssocL} constitute an
%% \vspace{.5mm}
iteration loop, which is initialized  (for $p \rmv=\rmv 0$) by
\begin{equation}
\zeta^{(0)}_{k \rightarrow m}\big(b^{(s)}_{n,m}\big) \ist= \sum_{a^{(s)}_{n,k}} \beta \big(a^{(s)}_{n,k}\big) 
  \ist \Psi\big(a^{(s)}_{n,k}\ist,b^{(s)}_{n,m}\big) \ist.
\label{eq:innerLoop}
\vspace{-.5mm}
\end{equation}
An efficient implementation of \eqref{eq:bp_iterativeDataAssoc_nu_1} and \eqref{eq:iterativeDataAssocL} is described in \cite{vontobel13} and \cite{williams14}. 
After the last iteration $p \rmv=\rmv P$, the messages
%% message $\eta\big(a^{(s)}_{n,k}\big)$ is calculated by multiplying the final results 
$\nu^{(P)}_{m \rightarrow k}\big(a^{(s)}_{n,k}\big)$, $m \in \cl{M}_n^{(s)}$ 
%% of the iteration,
%% iterative BP message passing loop  \eqref{eq:bp_iterativeDataAssoc_nu_1}, \eqref{eq:iterativeDataAssocL} are multiplied, 
are multiplied, 
\vspace{-.5mm}
i.e.,
\begin{equation}
\eta\big(a^{(s)}_{n,k}\big) \ist=  \prod_{m= 1}^{M^{(s)}_{n}}
%% \prod_{m \in \cl{M}_n^{(s)}} 
\rmv \nu^{(P)}_{m \rightarrow k}\big(a^{(s)}_{n,k}\big) \ist. 
\label{eq:iterativeDataAssocO}
\vspace{-.5mm}
\end{equation}

\item \emph{Measurement update}:
\begin{align}
\gamma^{(s)}(\bd{x}_{n,k},1) &\ist= \sum_{a^{(s)}_{n,k}} \upsilon\big( \bd{x}_{n,k} \ist, 1, a^{(s)}_{n,k} ; \bd{z}_{n}^{(s)}\big) \hspace{.2mm} \eta\big(a^{(s)}_{n,k}\big) \nn\\[-2.5mm]
&\label{eq:bp_measmessB}\\[-2mm]
\gamma^{(s)}_{n,k} &\ist=\ist \eta\big(a^{(s)}_{n,k} \!=\rmv 0\big)
%% \ist=\ist \eta^{(s)}_{n,k} 
\ist. \nn
%% \vspace{-.5mm}
\end{align}

\end{enumerate}

%% \newpage %%%%%%%

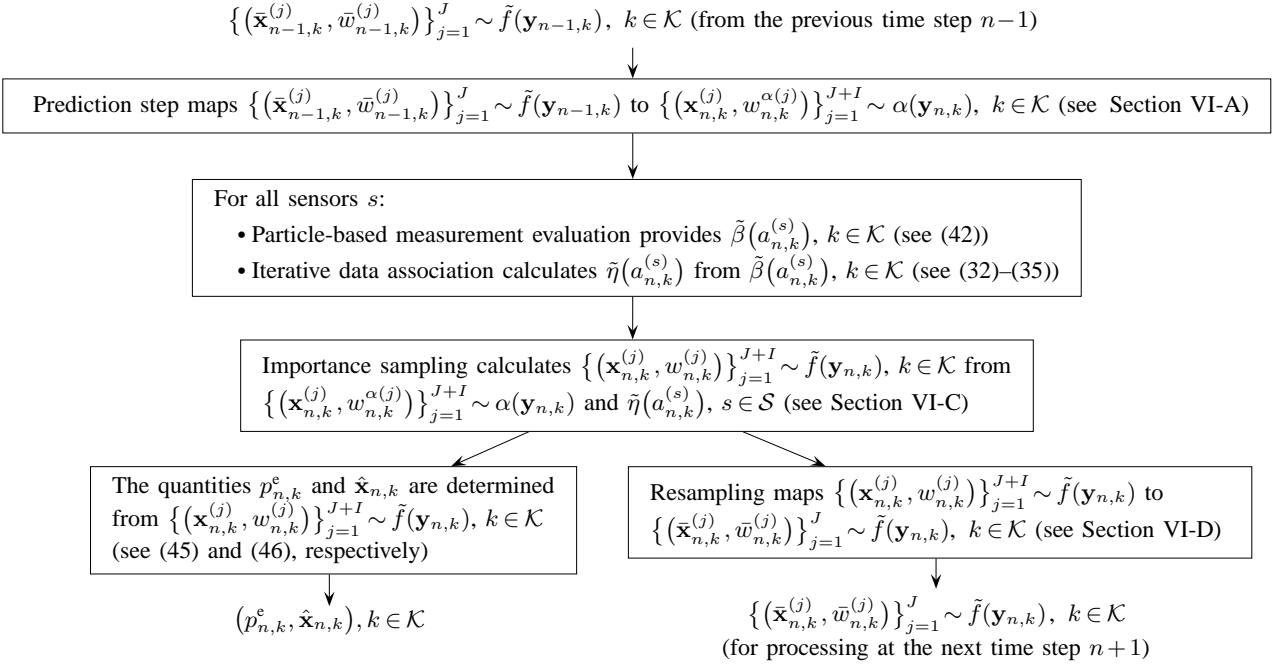
\begin{figure*}[t]
\vspace*{-2mm}
\centering \small \psset{xunit=7mm,yunit=7mm,runit=8mm}\psset{linewidth=0.3pt}
\begin{pspicture}(-3,1.8)(5.5,13.5)
\rput(0,13.1){\rnode{gb1}{ $\big\{ \big( \bar{\bd{x}}^{(j)}_{n-1,k} \ist, \bar{w}^{(j)}_{n-1,k} \big) \big\}_{j=1}^{J} \rmv\rmv\sim\rmv \tilde{f}(\bd{y}_{n-1,k})\ist$,\, 
$k \rmv\in\rmv \cl{K}$ (from the previous time step $n \!-\! 1$) }}

\rput(0,11.5){\rnode{gb1b}{ \psframebox{ $ \begin{array}{l}
\text{Prediction step maps $\big\{ \big( \bar{\bd{x}}^{(j)}_{n-1,k} \ist, \bar{w}^{(j)}_{n-1,k} \big) \big\}_{j=1}^{J} \rmv\rmv\sim\rmv \tilde{f}(\bd{y}_{n-1,k})$ to 
$\big\{ \big( \bd{x}^{(j)}_{n,k} \ist, w^{\alpha (j)}_{n,k} \big) \big\}_{j=1}^{J+I} \rmv\rmv\sim \alpha(\bd{y}_{n,k})$,\, $k \rmv\in\rmv \cl{K}$ (see\, Section \ref{sec:pred})}
\end{array}$}}}

\rput(0,9){\rnode{gb2o}{ \psframebox{$ \begin{array}{l}
\text{For all sensors $s$:}\\[.5mm]
 \hspace{3mm} \text{\tiny \raisebox{1.5pt}{$\bullet$} } \text{Particle-based measurement evaluation provides $\tilde{\beta}\big(a^{(s)}_{n,k}\big)$, $k \rmv\in\rmv \cl{K}$ (see \eqref{eq:approxBeta})} \\[.5mm]
  \hspace{3mm} \text{\tiny \raisebox{1.5pt}{$\bullet$} } \text{Iterative data association calculates $\tilde{\eta}\big(a^{(s)}_{n,k}\big)$ from $\tilde{\beta}\big(a^{(s)}_{n,k}\big)$, 
  $k \rmv\in\rmv \cl{K}$ (see \eqref{eq:bp_iterativeDataAssoc_nu_1}--\eqref{eq:iterativeDataAssocO})} 
  %% \\[1mm]
  %% \eqref{eq:iterativeDataAssocO}--\eqref{eq:innerLoop}
%% \text{ At this point, messages $\tilde{\eta}\big(a^{(s)}_{n,k}\big)$ are available for all PTs $k \rmv\in\rmv \cl{K}$ and all sensors $s \rmv\in\rmv \cl{S}$}
%% \{1,\dots,S \}$} 	
	\end{array}$}}}
	
\rput(0,6.2){\rnode{gb3o}{ \psframebox{$ \begin{array}{l}
\text{Importance sampling calculates $\big\{\big( \bd{x}^{(j)}_{n,k} \ist, w^{(j)}_{n,k} \big) \big\}_{j=1}^{J+I} \rmv\rmv\sim\rmv  \tilde{f}(\bd{y}_{n,k})$, 
$k \rmv\in\rmv \cl{K}$ from} \\ \text{$\big\{ \big( \bd{x}^{(j)}_{n,k} \ist, w^{\alpha (j)}_{n,k} \big) \big\}_{j=1}^{J+I} \rmv\rmv\sim\rmv \alpha(\bd{y}_{n,k})$ and $\tilde{\eta}\big(a^{(s)}_{n,k}\big)$, $s \rmv\in\rmv \cl{S}$
%% \{1,\dots,S \}$ 
(see Section \ref{sec:measurementUpdate}) }
\end{array}$}}}

\rput(5.7,3.8){\rnode{gb4a}{ \psframebox{$ \begin{array}{l}
\text{Resampling maps $\big\{ \big( \bd{x}^{(j)}_{n,k} \ist, w^{(j)}_{n,k} \big) \big\}_{j=1}^{J+I} \rmv\rmv\sim\rmv \tilde{f}(\bd{y}_{n,k})$ to} \\
\text{$\big\{ \big( \bar{\bd{x}}^{(j)}_{n,k} \ist, \bar{w}^{(j)}_{n,k} \big) \big\}_{j=1}^{J} \rmv\rmv\sim\rmv \tilde{f}(\bd{y}_{n,k}) $,\, $k \rmv\in\rmv \cl{K}$ (see Section \ref{sec:resampling}) }
\end{array}$}}}

\rput(-5.7,3.65){\rnode{gb4b}{ \psframebox{$ \begin{array}{l}
\text{The quantities $p^{\text{e}}_{n,k}$ and $\hat{\bd{x}}_{n,k}$ are determined} \\ 
\text{from $\big\{ \big( \bd{x}^{(j)}_{n,k} \ist, w^{(j)}_{n,k} \big) \big\}_{j=1}^{J+I} \rmv\rmv\sim\rmv  \tilde{f}(\bd{y}_{n,k})$,  $k \rmv\in\rmv \cl{K}$} \\ 
\text{(see \eqref{eq:approxExistProb_w} and \eqref{eq:approxStateEstimation}, respectively) }
\end{array}$}}}

\rput(5.7,1.65){\rnode{gb5a}{ $ \begin{array}{l} \\[-2.5mm] \text{\hspace{2.2mm} $\big\{ \big( \bar{\bd{x}}^{(j)}_{n,k} \ist, \bar{w}^{(j)}_{n,k} \big) \big\}_{j=1}^{J} \rmv\rmv\sim\rmv \tilde{f}(\bd{y}_{n,k})\ist$,\, $k \rmv\in\rmv \cl{K}$} \\[1mm] \text{(for processing at the next time step $n \rmv+\rmv 1$) }
\end{array}$}}
\rput(-5.7,1.75){\rnode{gb5b}{ $\big(p^{\text{e}}_{n,k},\hat{\bd{x}}_{n,k}\big), k \rmv\in\rmv \cl{K}$ }}
\ncline[linecolor=black,nodesepB=0mm,nodesepA=1mm,arrowsize=4pt]{->}{gb1}{gb1b}
\ncline[linecolor=black,nodesepB=0mm,nodesepA=0mm,arrowsize=4pt]{->}{gb1b}{gb2o}
\ncline[linecolor=black,nodesepB=0mm,nodesepA=0mm,arrowsize=4pt]{->}{gb2o}{gb3o}
\ncline[linecolor=black,nodesepB=0mm,nodesepA=0mm,arrowsize=4pt]{->}{gb3o}{gb4a}
\ncline[linecolor=black,nodesepB=0mm,nodesepA=0mm,arrowsize=4pt]{->}{gb3o}{gb4b}
\ncline[linecolor=black,nodesepB=0mm,nodesepA=0mm,arrowsize=4pt]{->}{gb4a}{gb5a}
\ncline[linecolor=black,nodesepB=0mm,nodesepA=0mm,arrowsize=4pt]{->}{gb4b}{gb5b}

\end{pspicture}
\vspace{6mm}
\renewcommand{\baselinestretch}{1.15}\small\normalsize
\caption{Flowchart of a particle-based implementation of the proposed BP method. The symbol $\sim$ expresses the fact that a set of particles and weights represents 
a certain distribution.} 
\label{fig:flowChart}
\vspace{1mm}
\end{figure*}

Finally, 
%% the 
\emph{beliefs} $\tilde{f}(\bd{y}_{n,k}) = \tilde{f}(\bd{x}_{n,k},r_{n,k})$ approximating the marginal posterior pdfs $f(\bd{y}_{n,k} | \bd{z} ) = f(\bd{x}_{n,k},r_{n,k} | \bd{z} )$
are obtained as the 
\vspace{-1mm}
products
\begin{align}
\tilde{f}(\bd{x}_{n,k},1) &\ist=\ist \frac{1}{C_{n,k}} \ist\ist \alpha(\bd{x}_{n,k},1) \prod_{s=1}^S
%% \prod_{s \in \cl{S}}
\gamma^{(s)}(\bd{x}_{n,k},1) \label{eq:belief_Q_1}\\[.5mm]
\tilde{f}_{n,k} &\ist=\ist \frac{1}{C_{n,k}} \ist\ist \alpha_{n,k} \prod_{s=1}^S
%% \prod_{s \in \cl{S}}
\gamma^{(s)}_{n,k} \,, \label{eq:belief_Q_0}\\[-5.5mm]
\nn
\end{align}
with the normalization constant 
%% \vspace{-1mm}
\begin{align}
C_{n,k} =\rmv \int \rmv \rmv \alpha(\bd{x}_{n,k},1) \prod_{s=1}^S \rmv\gamma^{(s)}(\bd{x}_{n,k},1) \ist \mathrm{d} \bd{x}_{n,k}
   \ist+\ist \alpha_{n,k} \rmv \prod_{s=1}^S \rmv\gamma^{(s)}_{n,k} \ist. \nn \\[-2mm]
  \label{eq:normalization} \\[-6.6mm]
  \nn
  %\vspace{-1.5mm}
\end{align}
Note that $f_{\text{D}}^2(\bd{x}_{n,k}) = f_{\text{D}}(\bd{x}_{n,k})$ (cf. Section \ref{sec:pot-targ_statist}) was used to obtain \eqref{eq:belief_Q_0}.
Because the belief $\tilde{f}(\bd{x}_{n,k},1)$ in \eqref{eq:belief_Q_1} 
\pagebreak %%%%%%%
approximates $f(\bd{x}_{n,k}, r_{n,k} \!=\! 1 |\bd{z})$, it can  
be substituted for 
$f(\bd{x}_{n,k}, r_{n,k} \!=\! 1 |\bd{z})$ in \eqref{eq:margpost_r} and \eqref{eq:margpost_x}, thus providing the basis for Bayesian target detection and state estimation
as discussed in Section \ref{sec:prob}. A particle-based implementation of the above BP method that avoids the explicit evaluation of integrals and message products 
will be presented in Section \ref{sec:particleBased}.

The ``data association'' iteration loop \eqref{eq:bp_iterativeDataAssoc_nu_1}--\eqref{eq:iterativeDataAssocO}
%% --\eqref{eq:iterativeDataAssocL} 
involves
%% is based 
solely messages related to discrete random variables.
%% , ???tackles the data association problem at each sensor in an efficient way. 
Being based on loopy BP, it does not perform an exact 
%% calculation of the 
marginalization \cite{kschischang01,wainwright08,loeliger}. 
%% in \eqref{eq:bp_extrinfo}). 
However, its high accuracy has been demonstrated numerically \cite{chertkov10, williams14} (see also Section \ref{sec:perf-comp}), 
and 
%% furthermore 
its convergence has been proven \cite{vontobel13, williams14}.

\vspace{-1mm}

\subsection{Scalability} %%?? Interpretation, Convergence, and Scalability
\label{sec:scalability}
%%%%%%%%%%%%%%%%%%%%%%%%%%%%%%%%%%%%%%%%%%%%%%%%%%%%%%%%%%%

The main
%% principal 
advantage of the 
%% proposed 
BP message passing method described in Section \ref{sec:severalsensors_bp} is its scalability.
%%  characteristic. 
Assuming a fixed number $P$ of message passing iterations, the computational complexity of calculating the (approximate) marginal posterior pdfs of all the 
%\pagebreak %%%%%%%%
target states is only linear in the number of sensors $S$ (see \eqref{eq:belief_Q_1}, \eqref{eq:belief_Q_0}). 
Moreover, the complexity of the operations \eqref{eq:bp_measevalution2}--\eqref{eq:bp_measmessB} performed for a given
%% each 
sensor $s \in \cl{S}$ 
scales as $\cl{O}\big(K M^{(s)}_n\big)$, where
%% . (Note that 
$M^{(s)}_n$ increases linearly 
%% is proportional to 
with the 
%% maximum 
number of PTs $K$ and with
%% to the expected 
the number of false alarms.
%%  $\mu^{(s)}\rmv$.)
Thus,
%% This means that 
the overall 
%% computational 
complexity of our algorithm scales linearly 
in the number of sensors
%% , linear in the expected number of false alarms $\mu^{(s)}$, 
and quadratically in the number of PTs.
%% (because each additional target potentially results in an additional measurement at each sensor).
Note that in practical implementations, measurement gating \cite{barShalom95} can be used to further improve scalability.

Such favorable scaling 
%% characteristic 
is a consequence of the 
%% use of the 
``detailed'' 
%% factor graph corresponding to the 
factorization \eqref{eq:factorOverall}. This factorization, in turn, is due to the redundant formulation of the joint state-association estimation task
%% problem 
in terms of both target-related and measurement-related association variables as described in Sections \ref{sec:pot-targ} and \ref{sec:prior}.  
Using this 
%% detailed 
factorization, an increase in the number of PTs, the number of sensors, or the number of measurements 
%\pagebreak %%%%%%%
leads to additional variable nodes 
in the factor graph (see Fig.\ \ref{fig:overallFG}) but not to higher dimensions of the messages passed between the nodes. 
The scalability of the proposed algorithm will be further analyzed in Section \ref{sec:sim-scal}.

%% \newpage %%%%%%%

%%%%%%%%%%%%%%%%%%%%%%%%%%%%%%%%%%%%%%%%%%%%%%%%%%%%%%%%%%%
\section{Particle-Based Implementation}
\label{sec:particleBased}
%%%%%%%%%%%%%%%%%%%%%%%%%%%%%%%%%%%%%%%%%%%%%%%%%%%%%%%%%%%

For general
%% arbitrary 
nonlinear and non-Gaussian measurement and state evolution models, 
%% some operations in the proposed BP scheme \eqref{eq:prediction}--\eqref{eq:belief_Q} are infeasible. More specifically, 
the integrals in \eqref{eq:mmse}, \eqref{eq:prediction_1}, \eqref{eq:prediction_0}, and \eqref{eq:bp_measevalution2} as well as the message products 
in \eqref{eq:belief_Q_1} and \eqref{eq:belief_Q_0} typically cannot be evaluated in closed form and are computationally infeasible. 
Therefore, we next present 
%% a particle-based scheme for 
an approximate particle-based implementation of these operations.
%% calculation of these expressions. 
In this implementation,
%% scheme, 
each belief $\tilde{f}(\bd{x}_{n,k}, r_{n,k})$ 
%% for each target $k$ and time $n$ are 
is represented by a set of particles and corresponding weights $\big\{ \big( \bar{\bd{x}}^{(j)}_{n,k} \ist, \bar{w}^{(j)}_{n,k} \big) \big\}_{j=1}^{J}$. 
%This set of weighted particles represents the dependency ???OF WHAT??? on both $\bd{x}_{n,k}$ and $r_{n,k}$. 
More specifically, $\tilde{f}(\bd{x}_{n,k},1)$ is represented by $\big\{ \big( \bar{\bd{x}}^{(j)}_{n,k} \ist, \bar{w}^{(j)}_{n,k} \big) \big\}_{j=1}^{J}$, and
$\tilde{f}(\bd{x}_{n,k},0)$ is 
%% then 
given implicitly by
%% due to 
the normalization property of $\tilde{f}(\bd{x}_{n,k},r_{n,k})$, i.e., $\tilde{f}(\bd{x}_{n,k},0) = 1 - \int \tilde{f}(\bd{x}_{n,k},1) \ist \mathrm{d} \bd{x}_{n,k}$. 
Contrary to conventional particle filtering \cite{arulampalam01,doucet01}, the particle weights $\bar{w}^{(j)}_{n,k}$, $j \in \{1,\dots,J\}$ do not sum to one;
instead, 
%% their sum approximates the probability of target existence, i.e., 
\be
p^{\sist\text{e}}_{n,k} \ist\triangleq\ist \sum^{J}_{j=1} \bar{w}^{(j)}_{n,k} \ist\approx\rmv \int \rmv \tilde{f}(\bd{x}_{n,k}, 1) \ist \text{d} \bd{x}_{n,k} \ist.
\label{eq:approxExistProb}
\ee
%% $\int \rmv \tilde{f}(\bd{x}_{n,k}, 1) \ist \text{d} \bd{x}_{n,k} \approx \sum^{J}_{j=1} \bar{w}^{(j)}_{n,k}$ 
%% and $\tilde{f}_{n,k}(0) \approx 1 - \sum^{J}_{j=1} \bar{w}^{(j)}_{n,k}$. 
Note that since $\int \rmv \tilde{f}(\bd{x}_{n,k}, 1) \ist \text{d} \bd{x}_{n,k}$ 
%% in turn 
approximates the 
%% marginal 
posterior probability of target existence $p(r_{n,k} \!=\! 1|\bd{z})$, it follows that the sum of weights $p^{\sist\text{e}}_{n,k}$ is approximately equal to $p(r_{n,k} \!=\! 1|\bd{z})$.

%% In the following, we will present particle-based implementations of \eqref{eq:prediction_1}, \eqref{eq:prediction_0}, \eqref{eq:bp_measevalution2}, 
%% and \eqref{eq:belief_Q_1}, \eqref{eq:belief_Q_0}. These 
The particle operations discussed in the remainder of this section are performed for all PTs $k \!\in\! \cl{K}$ in parallel. 
The resulting particle-based implementation of the overall BP method is summarized in Fig.\ \ref{fig:flowChart}.

\vspace{-1mm}

\subsection{Prediction}
\label{sec:pred}
%%%%%%%%%%%%%%%%%%%%%%%%%%%%%%%%%%%%%%%%%%%%%%%%%%%%%%%%%%%

For
%% At time 
$n \rmv\ge\! 1$ and for each PT $k \!\in\! \cl{K}$, $J$ particles and 
%% corresponding 
weights 
$\big\{ \big( \bar{\bd{x}}^{(j)}_{n-1,k} \ist, \bar{w}^{(j)}_{n-1,k} \!\rmv=\rmv p^{\sist\text{e}}_{n-1,k}/J \big) \big\}_{j=1}^{J}$ representing the 
previous belief $\tilde{f}( \bd{x}_{n-1,k}, r_{n-1,k})$ 
%% at the previous time $n \rmv-\! 1$ 
%% are available; these 
were calculated at the previous time $n \rmv-\! 1$ as described further below. 
%% Here, the weights $\bar{w}^{(j)}_{n-1,k}$, $j \in \{1,\dots,J\}$ are all equal and given by $p^{\sist\text{e}}_{n-1,k}/J$ (cf.\ Section \ref{sec:resampling}).
Weighted particles $\big\{ \big( \bd{x}^{(j)}_{n,k} \ist, w^{\alpha (j)}_{n,k} \big) \big\}_{j=1}^{J+I}$ representing the message  
$\alpha(\bd{x}_{n,k},1)$ in \eqref{eq:prediction_1} are now obtained as 
follows.\footnote{Note %%%%%%%%
that $\alpha_{n,k}$ in \eqref{eq:prediction_0} is again given implicitly by these weighted particles since $\alpha(\bd{x}_{n,k},r_{n,k})$ is 
normalized.} %%%%%%%%%%% 
First, for each particle $\bar{\bd{x}}^{(j)}_{n-1,k}\ist$, $j \in \{1,\dots,J\}$, one particle $\bd{x}^{(j)}_{n,k}$ is drawn 
from $f\big( \bd{x}_{n,k} \big| \bar{\bd{x}}^{(j)}_{n-1,k}\big)$. Next, $I$ additional ``birth particles'' $\big\{\bd{x}^{(j)}_{n,k}\big\}_{j=J+1}^{J+I}$ 
are drawn from the birth pdf $f_{\sist\text{b}}( \bd{x}_{n,k})$. 
Finally, weights $w^{\alpha(j)}_{n,k}$,
%%  for all particles 
$j \in \{1,\dots,J+I\}$ are obtained as
\begin{align}
\hspace{-1mm} w^{\alpha (j)}_{n,k} \rmv= \begin{cases} 
   p^{\sist\text{s}}_{n,k} \ist \bar{w}^{(j)}_{n-1,k} \ist, & \! j \in \{1, \dots, J \} \\[1mm]
   %% \frac{1}{I} \ist 
   p^{\sist\text{b}}_{n,k} (1 \rmv- p^{\sist\text{e}}_{n-1,k} ) / I \ist, & \! j \in \{J\!+\!1, \dots, J \!+\! I \} \ist. 
  \end{cases} \!\!
\label{eq:alphaweights} \\[-5.5mm]
\nn
\end{align}
Here, $p^{\sist\text{e}}_{n-1,k} = \sum^{J}_{j=1} \bar{w}^{(j)}_{n-1,k}$ (cf. \eqref{eq:approxExistProb}). 
Note that the proposal distribution \cite{arulampalam01,doucet01} underlying \eqref{eq:alphaweights} is 
%% here, as proposal distribution we chose 
$f\big( \bd{x}_{n,k} \big| \bd{x}^{(j)}_{n-1,k})$ for 
%% all particles 
$j \in \{1,\dots,J\}$ and $f_{\sist\text{b}}( \bd{x}_{n,k})$ for 
%% all particles 
$j \in \{J \rmv+\rmv 1,\dots,J \rmv+\rmv I\}$. A more general expression for particle-based prediction with an arbitrary proposal distribution can be found in \cite{ristic13}.

\vspace{-1mm}

\subsection{Measurement Evaluation}
%%%%%%%%%%%%%%%%%%%%%%%%%%%%%%%%%%%%%%%%%%%%%%%%%%%%%%%%%%%

For each sensor $s \!\in\! \cl{S}$, an approximation $\tilde{\beta}\big(a^{(s)}_{n,k}\big)$ of the message $\beta\big(a^{(s)}_{n,k}\big)$ 
in \eqref{eq:bp_measevalution2} can be calculated from the weighted particles $\big\{ \big( \bd{x}^{(j)}_{n,k} \ist, w^{\alpha (j)}_{n,k} \big) \big\}_{j=1}^{J+I}$ 
representing $\alpha(\bd{x}_{n,k}, r_{n,k})$ as
\begin{align}
\tilde{\beta}\big(a^{(s)}_{n,k}\big) &\ist= \sum^{J+I}_{j=1} \upsilon\big( \bd{x}^{(j)}_{n,k}, 1, a^{(s)}_{n,k}; \bd{z}_{n}^{(s)} \big) \ist w^{\alpha (j)}_{n,k} \nn\\[-2mm]
&\hspace{16mm}+\ist 1\big(a^{(s)}_{n,k}\big) \Bigg(\rmv 1-\sum^{J+I}_{j=1} w^{\alpha (j)}_{n,k}\rmv \Bigg) \ist. \label{eq:approxBeta}
\end{align}
Here, 
%% the first term, 
$\sum^{J+I}_{j=1} \upsilon\big( \bd{x}^{(j)}_{n,k}, 1, a^{(s)}_{n,k}; \bd{z}_{n}^{(s)} \big) \ist w^{\alpha (j)}_{n,k}$ 
provides a Monte Carlo 
%% integration 
approximation \cite{doucet01} of 
$\int \rmv \upsilon\big( \bd{x}_{n,k}, 1, a^{(s)}_{n,k}; \bd{z}_{n}^{(s)} \big) \ist \alpha(\bd{x}_{n,k}, 1)$\linebreak %%%%%%% \, 
$\times \ist \mathrm{d}\bd{x}_{n,k}$ in \eqref{eq:bp_measevalution2}, 
and 
%% the second term, 
$1\big(a^{(s)}_{n,k}\big) \big( 1-\sum^{J+I}_{j=1} w^{\alpha (j)}_{n,k} \big)$ 
provides an approximation of $1\big(a^{(s)}_{n,k}\big) \ist \alpha_{n,k}$ in \eqref{eq:bp_measevalution2}.
Note that
%% , similar to $p^{\sist\text{e}}_{n,k}$ in \eqref{eq:approxExistProb}, 
$\sum^{J+I}_{j=1} w^{\alpha (j)}_{n,k}$ can be interpreted as a ``predicted existence probability,''
and thus $1-\sum^{J+I}_{j=1} w^{\alpha (j)}_{n,k}$ can be interpreted as a ``predicted nonexistence probability,'' which approximates $\alpha_{n,k}$. 

\vspace{-1mm}

\subsection{Data Association, Measurement Update, Belief Calculation}
\label{sec:measurementUpdate}
%%%%%%%%%%%%%%%%%%%%%%%%%%%%%%%%%%%%%%%%%%%%%%%%%%%%%%%%%%%

\vspace{.4mm}

The approximate messages $\tilde{\beta}\big(a^{(s)}_{n,k}\big)$ obtained in \eqref{eq:approxBeta} are used in 
%% to initialize 
the data association loop,
%%  \eqref{eq:bp_iterativeDataAssoc_nu_1}--\eqref{eq:iterativeDataAssocO},
%% \eqref{eq:iterativeDataAssocO}--\eqref{eq:iterativeDataAssocL}, 
i.e., they are substituted for the messages 
$\beta\big(a^{(s)}_{n,k}\big)$ in \eqref{eq:iterativeDataAssocL} and \eqref{eq:innerLoop}.
%% to initialize the inner data association loop according to \eqref{eq:innerLoop}.
After convergence of the data association loop, approximate messages $\tilde{\eta}\big(a^{(s)}_{n,k}\big)$ 
(approximating the messages $\eta\big(a^{(s)}_{n,k}\big)$ in \eqref{eq:iterativeDataAssocO})
are then available for all PTs $k$ and all sensors $s$.
%% (These messages are approximations of $\eta_{\rightarrow}\big(a^{(s)}_{n,k}\big)$, since the inner data association loop 
%% was initialized with the approximate messages $\tilde{\beta}_{\rightarrow}\big(a^{(s)}_{n,k}\big)$ from \eqref{eq:approxBeta}.) 

%% We now implement 
Next, the measurement update step \eqref{eq:bp_measmessB} and the belief calculation step \eqref{eq:belief_Q_1}, \eqref{eq:belief_Q_0}
are implemented by means of importance sampling \cite{arulampalam01, doucet01}. To that end, we first rewrite the belief $\tilde{f}(\bd{x}_{n,k}, r_{n,k})$ in \eqref{eq:belief_Q_1}, \eqref{eq:belief_Q_0} 
by inserting 
%% the expression 
\eqref{eq:bp_measmessB},
%%  of the messages $\gamma^{(s)}(\bd{x}_{n,k}, r_{n,k})$, 
i.e.,
%% , $k \in \cl{K}$: 
%% from \eqref{eq:bp_measmess2} into \eqref{eq:belief_Q}
\begin{align}
\hspace{-.1mm}\tilde{f}(\bd{x}_{n,k},1) &\ist\propto\ist \alpha(\bd{x}_{n,k},1) \prod_{s=1}^S
%% \prod_{s \in \cl{S}}
\sum_{a^{(s)}_{n,k}} \upsilon\big( \bd{x}_{n,k} \ist, 1, a^{(s)}_{n,k} ; \bd{z}_{n}^{(s)}\big) \ist \tilde{\eta}\big(a^{(s)}_{n,k}\big) \nn\\[-4.5mm] 
\!\!\!\label{eq:approxbelief_1}\\[-1.5mm]
\tilde{f}_{n,k} &\ist\propto\ist \alpha_{n,k} \prod_{s=1}^S
%% \prod_{s \in \cl{S}} 
\tilde{\eta}\big(a^{(s)}_{n,k} \!=\rmv 0\big)
%% \tilde{\eta}\big(a^{(s)}_{n,k} \!=\! 0\big) 
\ist.
%% \big(a^{(s)}_{n,k}=0\big) \,,
\label{eq:approxbelief_0} \\[-5.5mm]
\nn
\end{align}
%% Note that 
Here, we also replaced $\eta\big(a^{(s)}_{n,k}\big)$ by its particle-based approximation 
$\tilde{\eta}\big(a^{(s)}_{n,k}\big)$, even though 
%% with some abuse of notation, 
we do not indicate this additional approximation in our notation $\tilde{f}(\bd{x}_{n,k}, r_{n,k})$. 
We now calculate nonnormalized weights corresponding to \eqref{eq:approxbelief_1} as
\begin{align*}
\hspace{-2mm}w^{\text{A}(j)}_{n,k} &=\ist w^{\alpha(j)}_{n,k} \prod_{s=1}^S
%% \prod_{s \in \cl{S}}
\sum_{a^{(s)}_{n,k}} \upsilon\big( \bd{x}^{(j)}_{n,k} \ist, 1, a^{(s)}_{n,k} ; \bd{z}_{n}^{(s)}\big) 
  \ist \tilde{\eta}\big(a^{(s)}_{n,k}\big) \ist, \nn \\[-3mm]
& \hspace{42mm}  j \in \{1,\dots,J+I\} \ist.
%% \label{eq:weights1}
\end{align*}
Note that this expression is based on
%% involves
%% corresponds to
%% Here, we used 
importance sampling with proposal density $\alpha(\bd{x}_{n,k},1)$ (represented by the weighted particles 
$\big\{ \big( \bd{x}^{(j)}_{n,k} \ist, w^{\alpha (j)}_{n,k} \big) \big\}_{j=1}^{J+I}$).
Similarly, we calculate a single nonnormalized weight corresponding to \eqref{eq:approxbelief_0} 
\vspace{.5mm}
as
\[
w^{\text{B}}_{n,k} =\ist \Bigg(\rmv 1-\sum^{J+I}_{j=1} w^{\alpha (j)}_{n,k}\rmv \Bigg) \prod_{s=1}^S
%% \prod_{s \in \cl{S}} 
\tilde{\eta}\big(a^{(s)}_{n,k} \!=\rmv 0\big) \ist ,
%% \label{eq:weights2}
%% \vspace{-.5mm}
\]
in which $1-\sum^{J+I}_{j=1} w^{\alpha(j)}_{n,k}\rmv$ approximates $\alpha_{n,k}$.

Next, weighted particles $\big\{ \big( \bd{x}^{(j)}_{n,k} \ist, w^{(j)}_{n,k} \big) \big\}_{j=1}^{J+I}$ representing the belief $\tilde{f}(\bd{x}_{n,k},r_{n,k})$ 
are obtained by using the particles $\big\{\bd{x}^{(j)}_{n,k}\big\}_{j=1}^{J+I}$ representing $\alpha(\bd{x}_{n,k},r_{n,k})$ and calculating the corresponding weights 
\vspace{-1.5mm}
as
\begin{equation}
w^{(j)}_{n,k} \ist=\ist \frac{w^{\text{A}(j)}_{n,k}}{w^{\text{B}}_{n,k}+\sum^{J+I}_{j'=1} \ist w^{\text{A}(j')}_{n,k}} \ist.
\nn
\end{equation}
Here, $w^{\text{B}}_{n,k}+\sum^{J+I}_{j=1} \ist w^{\text{A}(j)}_{n,k}$ is a particle-based approximation of the normalization constant $C_{n,k}$ in \eqref{eq:normalization}. We note that 
%% (cf.\ \eqref{eq:approxExistProb}) 
\be
p^{\sist\text{e}}_{n,k} \ist=\ist \sum^{J+I}_{j=1} w^{(j)}_{n,k} \ist.
\label{eq:approxExistProb_w}
\vspace{-1mm}
\ee

\subsection{Target Detection, State Estimation, Resampling}
\label{sec:resampling}
%%%%%%%%%%%%%%%%%%%%%%%%%%%%%%%%%%%%%%%%%%%%%%%%%%%%%%%%%%%

\vspace{.7mm}

The weighted particles $\big\{ \big( \bd{x}^{(j)}_{n,k} \ist, w^{(j)}_{n,k} \big) \big\}_{j=1}^{J+I}$ can now be used for target detection and estimation.
%%  as discussed in what follows. 
First, for each PT $k$, an approximation $p^{\sist\text{e}}_{n,k}$ of the existence probability $p(r_{n,k} \!=\! 1|\bd{z})$ is calculated  
from the particle weights $\big\{ w^{(j)}_{n,k} \big\}_{j=1}^{J+I}$ as in \eqref{eq:approxExistProb_w}. 
%% (with $J$ replaced by $J+I$). 
PT $k$ is then detected (i.e., considered to exist) if $p^{\sist\text{e}}_{n,k}$ is above a threshold $P_{\text{th}}$ (cf.\ Section \ref{sec:prob}).
For the detected targets $k$, an approximation of the MMSE state estimate $\hat{\bd{x}}^\text{MMSE}_{n,k}$ in \eqref{eq:mmse} is
%% can be 
calculated according 
\vspace{-.5mm}
to
\be
%% p^{\sist\text{e}}_{n,k} &= \sum_{j=1}^{J} \rmv w_{n,k}^{(j)}\,.\nn\\
\hat{\mathbf{x}}_{n,k} %%^{\text{MMSE}}
 \ist=\ist \frac{1}{p^{\sist\text{e}}_{n,k}} \sum_{j=1}^{J+I} \rmv w_{n,k}^{(j)} \ist \mathbf{x}_{n,k}^{(j)} \ist. 
\label{eq:approxStateEstimation}
%% \vspace{-.5mm}
\ee

Finally, as a preparation for the next time step $n+1$, a resampling step \cite{doucet01,arulampalam01} is performed to reduce the number of particles to $J$ and to avoid degeneracy effects. The resampling results in equally weighted particles $\big\{\bar{\bd{x}}^{(j)}_{n,k}\big\}_{j=1}^{J}$; the corresponding weights are given by $\bar{w}_{n,k}^{(j)} = \bar{w}_{n,k} = \frac{1}{J} \sum_{j'=1}^{I+J} w^{(j')}_{n,k}$, $j \in \{1, \dots, J \}$.

\vspace{-1mm}

%% \newpage %%%%%%

%%%%%%%%%%%%%%%%%%%%%%%%%%%%%%%%%%%%%%%%%%%%%%%%%%%%%%%%%%%
\section{Choice of Birth and Survival Parameters}
\label{sec:trackManagement}
%%%%%%%%%%%%%%%%%%%%%%%%%%%%%%%%%%%%%%%%%%%%%%%%%%%%%%%%%%%

We next present a scheme for choosing the birth pdfs $f_{\sist\text{b}}( \bd{x}_{n,k})$, birth probabilities $p_{n,k}^{\text{b}}$, and survival probabilities $p_{n,k}^{\text{s}}$
%%  for the potential targets $k \rmv\in\rmv \cl{K}$ at time $n$ 
(see Section \ref{sec:pot-targ_statist}). This scheme is heuristic but results in scalability with respect to the number of sensors $S$ and, as demonstrated in 
Section \ref{sec:perf-comp}, in good detection and tracking performance. 
%% guarantees that the proposed algorithm remains scalable in the number of sensors $S$; furthermore, as demonstrated in Section \ref{sec:simres}, 
%% it results in a good detection and tracking performance. 
It is based on the standard assumption that the number of newly born targets obeys a Poisson distribution with mean $\mu^\text{b}$, and existing targets survive with 
a fixed, specified probability $p^{\text{s}}$ \cite{mahler2007statistical}.  

We first distinguish between ``reliable'' and ``unreliable'' PTs at time $n \rmv-\! 1$ by comparing the PT existence probabilities 
$p^{\sist\text{e}}_{n-1,k}$ with a reliability threshold $R_{\text{th}}$: PT $k$ is considered reliable at time $n \rmv-\! 1$ if $p^{\sist\text{e}}_{n-1,k} \rmv>\rmv R_{\text{th}}$ and unreliable otherwise. 
Let $\cl{K}^{\text{r}}_{n-1}$ and $\cl{K}^{\text{u}}_{n-1}$ denote the sets of indices $k$ of reliable and unreliable PTs at time $n \rmv-\! 1$, respectively. 
For PTs 
%% that were reliable at time $n \rmv-\! 1$, i.e., 
$k \rmv\in\rmv \cl{K}^{\text{r}}_{n-1}$, we set the birth and survival probabilities at time $n$ to 
$p_{n,k}^{\text{b}} \rmv= 0$ and $p_{n,k}^{\text{s}} \rmv= p^{\text{s}}$, respectively. Since $p_{n,k}^{\text{b}} \rmv= 0$,
%%  for $k \in \cl{K}^{\text{r}}_{n-1}$, 
no birth pdf is needed at time $n$ (cf.\ \eqref{eq:singleTargetStateTrans_0}).

For 
%% those 
PTs 
%% that were unreliable at time $n \rmv-\! 1$, i.e., 
$k \rmv\in\rmv \cl{K}^{\text{u}}_{n-1}$, we set $p_{n,k}^{\text{s}} \rmv= 0$ and $p_{n,k}^{\text{b}} \rmv= \mu^{\text{b}}/|\cl{K}^{\text{u}}_{n-1}|$,\linebreak %%%%%%%%
and we construct
%% use a 
the birth pdf $f_{\sist\text{b}}( \bd{x}_{n,k})$ 
%% that is obtained 
as follows. Consider an arbitrary sensor $s_0$, and let 
%% $s_0$ denote the index of an arbitrary sensor and let 
$\cl{Z}^{(s_0)}_{n-1} \rmv\triangleq \big\{ \bd{z}^{(s_0)}_{n-1,m} \big\}_{m \in \cl{M}^{(s_0)}_{n-1}}$ denote the set of measurements of that sensor at time $n \rmv-\! 1$.
(Note that $\cl{Z}^{(s_0)}_{n-1}$ corresponds to the measurement vector $\bd{z}^{(s_0)}_{n-1} = \big[ \bd{z}^{(s_0)}_{n-1,m} ]_{m \in \cl{M}^{(s_0)}_{n-1}}$, 
with the difference that the elements $\bd{z}^{(s_0)}_{n-1,m}$ are considered 
%% ordered in $\bd{z}^{(s_0)}_{n-1}$ and 
unordered in $\cl{Z}^{(s_0)}_{n-1}$.) We partition $\cl{Z}^{(s_0)}_{n-1}$ into disjoint subsets 
$\cl{Z}^{\text{b}}_{n-1,k}$, $k \rmv\in\rmv \cl{K}^{\text{u}}_{n-1}$ such that 
%% $\bigcup_{k \in \cl{K}^{\text{u}}_{n-1}} \rmv \cl{Z}^{\text{b}}_{n-1,k} = \cl{Z}^{(s_0)}_{n-1}$ and 
the cardinalities of the $\cl{Z}^{\text{b}}_{n-1,k}$ differ at most by $1$, i.e., $\big| |\cl{Z}^{\text{b}}_{n-1,k}| - |\cl{Z}^{\text{b}}_{n-1,l}| \big| \le 1$ for any $k,l \rmv\in\rmv \cl{K}^{\text{u}}_{n-1}$.
Then, based on the $k$th measurement set $\cl{Z}^{\text{b}}_{n-1,k}$, we construct a corresponding ``adaptive birth pdf'' as
\[
f_{\sist\text{b}}( \bd{x}_{n,k}) \ist\triangleq\rmv \int \! f( \bd{x}_{n,k} | \bd{x}_{n-1,k}) \ist f_{\sist\text{b}}( \bd{x}_{n-1,k};\cl{Z}^{\text{b}}_{n-1,k})  
  \ist \mathrm{d} \bd{x}_{n-1,k} \ist,
\vspace{-.4mm}
\]
where the pdf $f_{\sist\text{b}}( \bd{x}_{n-1,k};\cl{Z}^{\text{b}}_{n-1,k})$ is
%% can be 
constructed using 
%% the measurements 
$\cl{Z}^{\text{b}}_{n-1,k}$ and prior knowledge (e.g., about the target velocity) as discussed in \cite{ristic12}. 
%% Finally, 
Particles representing $f_{\sist\text{b}}( \bd{x}_{n,k})$ are obtained by 
%% first 
drawing particles from $f_{\sist\text{b}}( \bd{x}_{n-1,k};\cl{Z}^{\text{b}}_{n-1,k})$ and performing particle-based prediction \cite{doucet01}.
%% , to obtain particles representing $f_{\sist\text{b}}( \bd{x}_{n,k})$. 

\vspace{-.5mm}

%%%%%%%%%%%%%%%%%%%%%%%%%%%%%%%%%%%%%%%%%%%%%%%%%%%%%%%%%%%
\section{Relation to Existing Methods}
\label{sec:existing methods}
%%%%%%%%%%%%%%%%%%%%%%%%%%%%%%%%%%%%%%%%%%%%%%%%%%%%%%%%%%%

%% Although the proposed algorithm was derived and formulated using the BP framework, which was not used previously for it 
Several
%% Certain 
aspects of the proposed method are related to existing methods, as discussed next.
%% in what follows. 

\begin{itemize}

\item The hybrid model for data association using both target-oriented and measurement-oriented association variables was previously proposed 
in \cite{chertkov10} and \cite{williams14}. In \cite{chertkov10}, BP
%% -based inference 
is used to estimate optical flow parameters.
%%  from an image sequence. 
In \cite{williams14}, BP is 
%% only 
used for 
%% approximate 
data association (without multitarget tracking). 

\item Our
%% The 
model for target existence 
%% used in our algorithm 
was previously used by the search-initialize-track filter in \cite{horridge09}, which, however, is not BP-based, considers only a single sensor, does not employ the 
%% efficient 
hybrid model for data association, and uses a different track initialization scheme. 

\item In the case of a single target and a single sensor, the proposed method reduces to the particle-based implementation of the Bernoulli filter \cite{ristic13} (which is derived using the FISST framework).

\item  The TOMB/P filter \cite{williams2015marg}, which is effectively a FISST-based variant of the JIPDA filter that uses BP 
and the hybrid data association model, 
%% from \cite{chertkov10} for data association, 
differs from the proposed method in the following respects: it is restricted to a single sensor and to linear-Gaussian state evolution and measurement models---see \cite{kropfreiter16} for an extension to nonlinear, non-Gaussian models---and the number of PTs (tracks) varies over time. 

%% \newpage %%%%%%%

\item  If the parameters of the proposed method are chosen such that all targets exist at all times (this special case 
was mentioned in Section \ref{sec:pot-targ_statist}, and was considered in our previous work in \cite{meyer15scalable}), then the method becomes similar to the 
Monte Carlo JPDA filter \cite{vermaak05} in that it uses 
a similar particle-based processing scheme. However, contrary to the Monte Carlo JPDA filter, the proposed method performs data association by means of BP, is based on the hybrid model for data association, and can also be used when the number of targets is unknown.

\end{itemize}

\section{Simulation Results}
\label{sec:simres}

%% In this section, we 
Next, we report 
%% and discuss 
simulation results assessing the performance of our method and comparing it with that of five previously proposed methods for multisensor-multi\-target tracking. 
%% and we study how certain parameters affect the performance.
%% mean OSPA (MOSPA) performance of the algorithms.

\vspace{-1mm}

%% \newpage %%%%%%%

\subsection{Simulation Setting}

We simulated up to five actual targets whose states consist of two-dimensional (2D) position and velocity, i.e., 
$\bd{x}_{n,k} \rmv= [x_{1,n,k} \;\ist x_{2,n,k} \;\ist \dot{x}_{1,n,k} \;\ist \dot{x}_{2,n,k}]^{\text{T}}\rmv$.
The targets move in a region of interest (ROI) given by $[-3000, \ist 3000] \times [-3000, \ist 3000]$ according to the constant-velocity motion model, i.e.,
$\bd{x}_{n,k} = \bd{A}\ist\bd{x}_{n-1,k} + \bd{W}\bd{u}_{n,k}$, 
%\pagebreak %%%%%%
where 
%% the matrices 
$\bd{A} \rmv\in\rmv \mathbb{R}^{4 \times 4}$ and $\bd{W} \rmv\in\rmv \mathbb{R}^{4 \times 2}$ are chosen as in \cite{kotecha03} and 
$\bd{u}_{n,k} \rmv\sim \cl{N}(\bd{0},\sigma^2_u \ist \bd{I}_2)$ with $\sigma^2_u \!=\rmv 0.025$ is an independent and identically distributed (iid) 
sequence of 2D Gaussian random vectors. The birth distribution $f_{\text{b}}(\bd{x}_{n,k})$, the birth probabilities $p_{n,k}^{\sist\text{b}}$, and the survival probabilities $p_{n,k}^{\sist\text{s}}$ were chosen as described in Section \ref{sec:trackManagement}, using the global birth probability $p^{\sist\text{b}} \rmv=\rmv 0.01$ and the global survival probability $p^{\sist\text{s}} \rmv=\rmv 0.999$. The number of PTs was set to $K \rmv=\rmv 8$.
%% To show the benefits of the proposed method and of the use of multiple sensors, we 
We considered a 
%% very 
challenging 
%% multitarget tracking 
scenario where the five target trajectories intersect 
%% cross 
at the ROI center. 
%% To obtain this behavior, 
The target trajectories were generated by first assuming that the 
five targets start from initial positions 
%% (at time $n=0$)  
uniformly placed on a circle of radius 1000 and
%% they 
move with an initial speed of 10 toward the ROI center, and then 
%% obtaining
%% From these preliminary target trajectories, 
%% the actual trajectories by 
letting the targets start to exist at times $n \rmv=\rmv 5$, $10$, $15$, $20$, and $25$.
%% , respectively.

%% In the simulated scenario $S$ sensors with 
The sensors are located uniformly 
%% placed 
on a circle of radius 3000 and perform range and bearing measurements within a measurement range of 6000. 
More specifically, within the measurement range, the target-generated measurements are given
by
\[
\bd{z}^{(s)}_{n,m} =\ist \big[ \, \big\| \tilde{\bd{x}}_{n,k} \!-\rmv \bd{p}^{(s)} \rmv\big\|  \;\,\ist \varphi\big(\tilde{\bd{x}}_{n,k},\bd{p}^{(s)}\big) \ist\big]^{\text{T}} \!+\ist \bd{v}^{(s)}_{n,m} \,, 
%% \label{eq:measModelSim}
\vspace{.5mm}
\]
where $\tilde{\bd{x}}_{n,k} \rmv\triangleq [x_{1,n,k} \;\ist x_{2,n,k}]^{\text{T}}\rmv$,
%%  is the ``position-related'' part of the target state $\bd{x}_{n,k}$, 
$\bd{p}^{(s)}$ is the position of sensor $s$,
$\varphi\big(\tilde{\bd{x}}_{n,k},\bd{p}^{(s)}\big)$ is the angle (in degrees) of the vector $\tilde{\bd{x}}_{n,k}$ relative to the vector $\bd{p}^{(s)}\rmv$, 
and $\bd{v}^{(s)}_{n,m} \sim \cl{N}(\bd{0},\bd{C}_v)$ with 
%% covariance matrix 
$\bd{C}_v = \mathrm{diag}\{10^2, 0.5^2\}$ is an iid sequence of 2D Gaussian random
%% measurement noise 
vectors. 
%% For all simulations the 
The false alarm pdf $f_{\text{FA}}\big( \bd{z}^{(s)}_{n,m} \big)$ is linearly increasing on $[0, 6000]$ and zero outside $[0, 6000]$ with respect to the range component, 
and uniform on $[0^{\circ},360^{\circ})$ with respect to the angle component. 
In Cartesian coordinates, this corresponds to a uniform distribution on the sensor's measurement area. 
%% The sensors are located at positions $\bd{p}^{(s)}$ that are uniformly placed on a circle of radius 3000 
%% and can perform measurements within an range of 6000. 
The mean number of false alarm measurements is $\mu^{(s)} \!\rmv=\! 2$ if not noted otherwise.

Our implementation of the proposed method used $J \rmv=\rmv 3000$ particles and $I \rmv=\rmv 3000$ birth particles for each PT. We performed $P \!=\rmv 20$ BP iterations for iterative data association. The threshold for target detection was $P_{\text{th}} \rmv=\rmv 0.5$, and the 
reliability threshold was $R_{\text{th}} \rmv=\rmv 10^{-3}\rmv$. We simulated 150 time steps $n$.

%% \vspace{1mm}

%% \subsection{Reference Methods and Simulation Parameters}
%% \newpage %%%%%%%

\vspace{-1mm}

\subsection{Performance Comparison}
\label{sec:perf-comp}

We compare the proposed BP method
%% method 
with particle implementations of the IC-PHD filter \cite{vo05,mahler2007statistical,nagappa11}, the IC-CPHD filter \cite{vo07,mahler2007statistical,nagappa11}, 
the IC-MB filter \cite{mahler2007statistical,vo09}, and the 
%% recently introduced 
partition-based MS-PHD and MS-CPHD filters \cite{nannuru15,nannuru15journal}. 
The ``IC-'' filters are straightforward
%% simple 
multisensor extensions 
%% of a corresponding single-sensor filter, 
performing a single-sensor update step sequentially for each sensor \cite{pao95,mahler2007statistical,nagappa11}.
The partition-based MS-(C)PHD filters approximate the 
%% true 
exact multisensor (C)PHD filters; they 
%% avoid the potentially poor performance of IC implementations and 
can outperform the ``IC-'' filters but have a higher computational complexity. Since the trellis algorithm used for partition extraction in the original formulation of the MS-(C)PHD filter \cite{nannuru15,nannuru15journal} is only suitable for a Gaussian mixture implementation of the filter, it was adapted to a particle-based implementation. 
We note that the exact multisensor (C)PHD filters are not computationally feasible for the simulated scenario since their complexity scales exponentially 
in the number of sensors and in the number of measurements per sensor \cite{mahler09a,mahler09b,delande11}. 
The performance of the various methods is measured by the Euclidean distance based OSPA metric with a cutoff parameter of 200 \cite{schuhmacher08}.
%%  whose calculation is based on the Euclidean distance. 

The (C)PHD-type filters use 24000 particles to represent the PHD
%% pdf 
of the target states, and they perform kmeans++ clustering \cite{gan07} for state estimation. 
The IC-MB filter uses 3000 particles to represent each Bernoulli component.
The maximum numbers of subsets and partitions used by the MS-(C)PHD filter are 120 and 720, respectively, similarly to \cite{nannuru15,nannuru15journal}. 
With the above-mentioned parameters, the runtime per time scan for a MATLAB implementation without gating
%%  of our method 
on a single core of an Intel Xeon E5-2640 v3 CPU was measured as 
0.07{\ist}s
%%  seconds 
for the proposed method, 
0.11{\ist}s for the IC-PHD filter, 
0.14{\ist}s for the IC-CPHD filter, 
0.27{\ist}s for the IC-MB filter,
13.21{\ist}s for the MS-PHD filter, and 
13.82{\ist}s for MS-CPHD filter. 
The high runtimes of the MS-(C)PHD
%% MS-PHD and MS-CPHD 
filters are due to the fact that in our particle-based implementation, the trellis algorithm used for partition extraction is computationally intensive.
%% very inefficient. 
We note that the efficient extraction of high-quality partitions in a particle-based implementation of the MS-(C)PHD filter is still an open 
%% research 
problem.
%At each time $n$, the birth intensities or pdfs of all methods were generated by using the measurements $\bd{z}^{(s')}_{n-1,m}$ from time $n \rmv-\! 1$ 
%of an arbitrarily chosen sensor $s'\rmv$ and a prior distribution of the target velocity (see \cite{meyer16scalable_ext} for details).
%with components $f_{\text{c}}(\bd{x}_{n,k}) = \int f( \bd{x}_{n,k} | \bd{x}_{n-1,k}) \iist C^{(s')}_{n-1,m} \iist f\big(\bd{z}^{(s')}_{n-1,m}\big|\tilde{\bd{x}}_{n-1,k}\big)$ $f_{\text{v}}\big(\dot{x}_{1,n-1,k},\dot{x}_{2,n-1,k}\big) \ist \mathrm{d}\bd{x}_{n-1,k}$, where $f\big(\bd{z}^{(s')}_{n-1,m}\big|\tilde{\bd{x}}_{n-1,k}\big)$ is the likelihood function (evaluated for measurement $\bd{z}^{(s')}_{n-1,m}$) corresponding to \eqref{eq:measModelSim}, $C^{(s')}_{n-1,m} \triangleq 1/\int f\big(\bd{z}^{(s')}_{n-1,m}\big|\tilde{\bd{x}}_{n-1,k}\big) \mathrm{d} \tilde{\bd{x}}_{n-1,k}$ is a normalization constant, and $f_{\text{v}}\big(\dot{x}_{1,n,k},\dot{x}_{2,n,k}\big) = \mathcal{N}(\bm{\mu^{(s)}}_{\mathrm{v}},\mathbf{C}_{\mathrm{v}} )$ with $\bm{\mu^{(s)}}_{\mathrm{v}} = [\ist 0 \;\, 0 \ist]^{\mathrm{T}}$ and $\mathbf{C}_{\mathrm{v}} \rmv= \mathrm{diag}\ist\{10^2, 10^2\}$ represents an assumed prior statistical knowledge about the target velocity $[\dot{x}_{1,n,k} \;\, \dot{x}_{2,n,k}]^{\mathrm{T}}$.

\begin{figure}
\vspace{.5mm}
\centering
\psfrag{s04}[t][t][0.8]{\color[rgb]{0,0,0}\setlength{\tabcolsep}{0pt}\begin{tabular}{c}\raisebox{-1mm}{$n$}\end{tabular}}
\psfrag{s03}[b][b][0.8]{\color[rgb]{0,0,0}\setlength{\tabcolsep}{0pt}\begin{tabular}{c}\raisebox{1mm}{MOSPA}\end{tabular}}
\psfrag{BP}[l][l][0.69]{\color[rgb]{0,0,0}BP (proposed)}
\psfrag{MS-CPHD}[l][l][0.69]{\color[rgb]{0,0,0}MS-CPHD}
\psfrag{IC-CPHD}[l][l][0.69]{\color[rgb]{0,0,0}IC-CPHD}
\psfrag{MS-PHD}[l][l][0.69]{\color[rgb]{0,0,0}MS-PHD}
\psfrag{IC-PHD}[l][l][0.69]{\color[rgb]{0,0,0}IC-PHD}
\psfrag{IC-MeMBer}[l][l][0.69]{\color[rgb]{0,0,0}IC-MB}
\psfrag{s10}[][]{\color[rgb]{0,0,0}\setlength{\tabcolsep}{0pt}\begin{tabular}{c} \end{tabular}}
\psfrag{s11}[][]{\color[rgb]{0,0,0}\setlength{\tabcolsep}{0pt}\begin{tabular}{c} \end{tabular}}
\psfrag{x01}[t][t][0.70]{150}
\psfrag{x02}[t][t][0.70]{25}
\psfrag{x03}[t][t][0.70]{50}
\psfrag{x04}[t][t][0.70]{75}
\psfrag{x05}[t][t][0.70]{100}
\psfrag{x06}[t][t][0.70]{125}
\psfrag{v01}[r][r][0.70]{0}
\psfrag{v02}[r][r][0.70]{40}
\psfrag{v03}[r][r][0.70]{80}
\psfrag{v04}[r][r][0.70]{120}
\psfrag{v05}[r][r][0.70]{160}
\psfrag{v06}[r][r][0.70]{200}
\includegraphics[height=50mm, width=70mm]{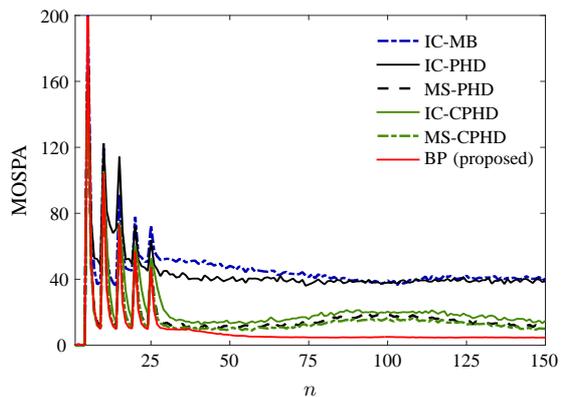}
\vspace{-.5mm}
\renewcommand{\baselinestretch}{1.05}\small\normalsize
\caption{MOSPA error 
%% of six multisensor-multitarget tracking algorithms 
versus time $n$ for $S \rmv=\! 3$ sensors,  
%% detection probabil\-ity 
$P_\text{d} \!=\! 0.8$, and $\mu^{(s)} \!\rmv=\! 2$.}
\label{fig:OSPAvsTime}
%% \vspace{-1mm}
\end{figure}

Fig.\ \ref{fig:OSPAvsTime} shows the mean OSPA (MOSPA) error---averaged over 400 simulation runs---of all
%% the various 
methods versus time $n$, assuming $S \rmv=\! 3$ sensors with a 
%% sensor 
detection probability of $P^{(s)}_{\text{d}}(\bd{x}_{n,k}) \rmv=\rmv P_\text{d} \!=\rmv 0.8$. The error exhibits peaks at times $n \rmv=\rmv 5$, $10$, $15$, $20$, and $25$ because of
%% related to 
target births.
%%  as discussed above. 
However, very soon after a target birth, the proposed method as well as the IC-CPHD, MS-PHD, and MS-CPHD filters are able to reliably estimate the 
%% correct 
number of targets. 
%% Furthermore, the 
The proposed method is seen to outperform all the other methods. 
In particular, it outperforms the IC-CPHD, MS-PHD, and MS-CPHD filters mainly because particle implementations of (C)PHD filters
involve a potentially unreliable clustering step.  
%%  for state extraction. 
This clustering step is especially unreliable
%% prone to error 
for targets that are close to each other. This fact explains the higher MOSPA error of the IC-CPHD, MS-PHD, and MS-CPHD 
filters around $n=100$, i.e., around the time when
%% during the period where 
the target trajectories intersect in the ROI center.
Finally, the MOSPA error of the IC-PHD and IC-MB filters is seen to be significantly larger than that of the other methods; this is caused by the inability of
these filters to reliably estimate the number of targets. 
We note that for sensors with different probabilities of detection,
%% in scenarios where the probability of detection is different among sensors, 
the performance loss of IC-(C)PHD filters relative to MS-(C)PHD filters tends to be larger than in Fig.\ \ref{fig:OSPAvsTime}
\cite{nannuru15,nannuru15journal}.

\begin{figure}
%% \vspace{3.5mm}
\centering
\psfrag{s03}[t][t][0.8]{\color[rgb]{0,0,0}\setlength{\tabcolsep}{0pt}\begin{tabular}{c}\raisebox{-1mm}{$P_{\text{d}}$}\end{tabular}}
\psfrag{s04}[b][b][0.8]{\color[rgb]{0,0,0}\setlength{\tabcolsep}{0pt}\begin{tabular}{c}\raisebox{0.3mm}{MOSPA}\end{tabular}}
\psfrag{BP}[l][l][0.69]{\color[rgb]{0,0,0}BP (proposed)}
\psfrag{MS-CPHD}[l][l][0.69]{\color[rgb]{0,0,0}MS-CPHD}
\psfrag{IC-CPHD}[l][l][0.69]{\color[rgb]{0,0,0}IC-CPHD}
\psfrag{MS-PHD}[l][l][0.69]{\color[rgb]{0,0,0}MS-PHD}
\psfrag{IC-PHD}[l][l][0.69]{\color[rgb]{0,0,0}IC-PHD}
\psfrag{IC-MeMBer}[l][l][0.69]{\color[rgb]{0,0,0}IC-MB}
\psfrag{s10}[][]{\color[rgb]{0,0,0}\setlength{\tabcolsep}{0pt}\begin{tabular}{c} \end{tabular}}
\psfrag{s11}[][]{\color[rgb]{0,0,0}\setlength{\tabcolsep}{0pt}\begin{tabular}{c} \end{tabular}}
\psfrag{x01}[t][t][0.70]{0.2}
\psfrag{x02}[t][t][0.70]{0.3}
\psfrag{x03}[t][t][0.70]{0.4}
\psfrag{x04}[t][t][0.70]{0.5}
\psfrag{x05}[t][t][0.70]{0.6}
\psfrag{x06}[t][t][0.70]{0.7}
\psfrag{x07}[t][t][0.70]{0.8}
\psfrag{x08}[t][t][0.70]{0.9}
\psfrag{v01}[r][r][0.70]{0}
\psfrag{v02}[r][r][0.70]{20}
\psfrag{v03}[r][r][0.70]{40}
\psfrag{v04}[r][r][0.70]{60}
\psfrag{v05}[r][r][0.70]{80}
\psfrag{v06}[r][r][0.70]{100}
\psfrag{v07}[r][r][0.70]{120}
\includegraphics[height=50mm, width=70mm]{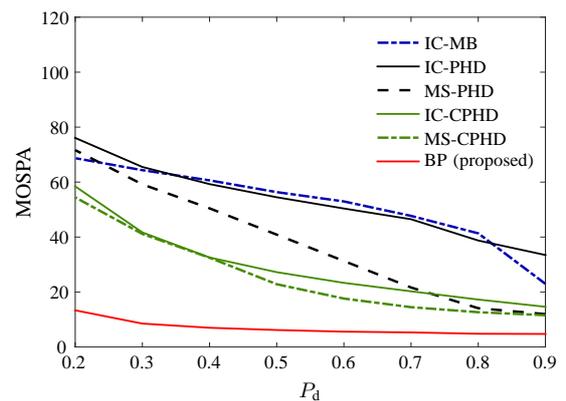}
\vspace{-.5mm}
\renewcommand{\baselinestretch}{1.05}\small\normalsize
\caption{Time-averaged MOSPA error 
%% of six multisensor-multitarget tracking algorithms 
versus 
%% detection probabil\-ity 
$P_\text{d}$ for $S \rmv=\! 3$ sensors and $\mu^{(s)} \!\rmv=\! 2$.}
\vspace{-5mm}
\label{fig:OSPAvsPD}
\vspace{1.5mm}
\end{figure}

\begin{figure}
\vspace{-1mm}
\centering
\psfrag{s01}[t][t][0.8]{\color[rgb]{0,0,0}\setlength{\tabcolsep}{0pt}\begin{tabular}{c}\raisebox{-1mm}{$S$}\end{tabular}}
\psfrag{s02}[b][b][0.8]{\color[rgb]{0,0,0}\setlength{\tabcolsep}{0pt}\begin{tabular}{c}\raisebox{1mm}{MOSPA}\end{tabular}}
\psfrag{BP}[l][l][0.69]{\color[rgb]{0,0,0}BP (proposed)}
\psfrag{MS-CPHD}[l][l][0.69]{\color[rgb]{0,0,0}MS-CPHD}
\psfrag{IC-CPHD}[l][l][0.69]{\color[rgb]{0,0,0}IC-CPHD}
\psfrag{MS-PHD}[l][l][0.69]{\color[rgb]{0,0,0}MS-PHD}
\psfrag{IC-PHD}[l][l][0.69]{\color[rgb]{0,0,0}IC-PHD}
\psfrag{IC-MeMBer}[l][l][0.69]{\color[rgb]{0,0,0}IC-MB}
\psfrag{s10}[][]{\color[rgb]{0,0,0}\setlength{\tabcolsep}{0pt}\begin{tabular}{c} \end{tabular}}
\psfrag{s11}[][]{\color[rgb]{0,0,0}\setlength{\tabcolsep}{0pt}\begin{tabular}{c} \end{tabular}}
\psfrag{x01}[t][t][0.70]{1}
\psfrag{x02}[t][t][0.70]{2}
\psfrag{x03}[t][t][0.70]{3}
\psfrag{x04}[t][t][0.70]{4}
\psfrag{x05}[t][t][0.70]{5}
\psfrag{x06}[t][t][0.70]{6}
\psfrag{x07}[t][t][0.70]{7}
\psfrag{x08}[t][t][0.70]{8}
\psfrag{v01}[r][r][0.70]{0}
\psfrag{v02}[r][r][0.70]{20}
\psfrag{v03}[r][r][0.70]{40}
\psfrag{v04}[r][r][0.70]{60}
\psfrag{v05}[r][r][0.70]{80}
\psfrag{v06}[r][r][0.70]{100}
\psfrag{v07}[r][r][0.70]{120}
\includegraphics[height=50mm, width=70mm]{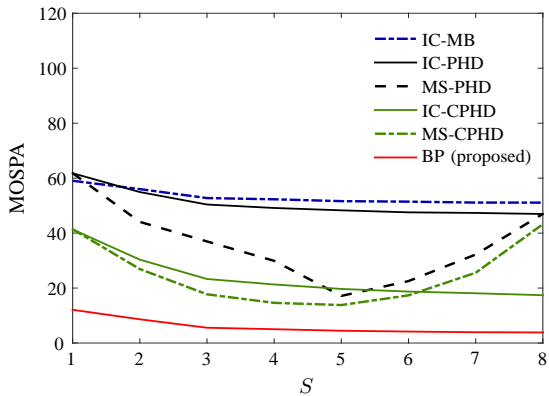}
\vspace{-.5mm}
\renewcommand{\baselinestretch}{1.05}\small\normalsize
\caption{Time-averaged MOSPA error 
%% of six multisensor-multitarget tracking algorithms 
versus number of sensors $S$ for 
%% detection probabil\-ity 
$P_\text{d} \!=\! 0.6$ and $\mu^{(s)} \!\rmv=\! 2$.}
\label{fig:OSPAvsSensors}
%% \vspace{-1.5mm}
\end{figure}
%% We performed 1000 Monte Carlo iterations.

%% \newpage %%%%%%%

\begin{figure}
\centering
\psfrag{s01}[b][b][0.8]{\color[rgb]{0,0,0}\setlength{\tabcolsep}{0pt}\begin{tabular}{c}\raisebox{1mm}{MOSPA}\end{tabular}}
\psfrag{s03}[t][t][0.8]{\raisebox{-3mm}{$\mu^{(s)}$}}
\psfrag{BP}[l][l][0.69]{\color[rgb]{0,0,0}BP (proposed)}
\psfrag{MS-CPHD}[l][l][0.69]{\color[rgb]{0,0,0}MS-CPHD}
\psfrag{IC-CPHD}[l][l][0.69]{\color[rgb]{0,0,0}IC-CPHD}
\psfrag{MS-PHD}[l][l][0.69]{\color[rgb]{0,0,0}MS-PHD}
\psfrag{IC-PHD}[l][l][0.69]{\color[rgb]{0,0,0}IC-PHD}
\psfrag{IC-MeMBer}[l][l][0.69]{\color[rgb]{0,0,0}IC-MB}
\psfrag{s10}[][]{\color[rgb]{0,0,0}\setlength{\tabcolsep}{0pt}\begin{tabular}{c} \end{tabular}}
\psfrag{s11}[][]{\color[rgb]{0,0,0}\setlength{\tabcolsep}{0pt}\begin{tabular}{c} \end{tabular}}
\psfrag{x01}[t][t][0.70]{1}
\psfrag{x02}[t][t][0.70]{10}
\psfrag{x03}[t][t][0.70]{20}
\psfrag{x04}[t][t][0.70]{30}
\psfrag{x05}[t][t][0.70]{40}
\psfrag{x06}[t][t][0.70]{50}
\psfrag{v01}[r][r][0.70]{0}
\psfrag{v02}[r][r][0.70]{20}
\psfrag{v03}[r][r][0.70]{40}
\psfrag{v04}[r][r][0.70]{60}
\psfrag{v05}[r][r][0.70]{80}
\psfrag{v06}[r][r][0.70]{100}
\psfrag{v07}[r][r][0.70]{120}
\includegraphics[height=50mm, width=70mm]{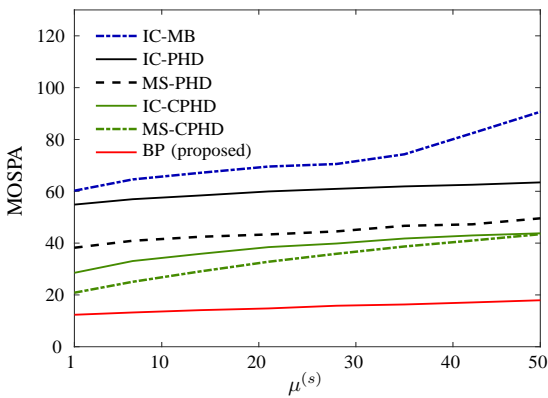}
\vspace{-.5mm}
\renewcommand{\baselinestretch}{1.05}\small\normalsize
\caption{Time-averaged MOSPA error versus mean number of false alarms $\mu^{(s)}$ for $P_\text{d} \!=\! 0.6$ and $S \!=\! 3$.}
\label{fig:OSPAvsClutter}
%% \vspace{-1.5mm}
\end{figure}

Fig.\ \ref{fig:OSPAvsPD} shows the time-averaged MOSPA error---averaged over 
%% the 
time steps $n \rmv\in\rmv \{50,\dots,150\}$---versus the detection probability $P_\text{d}$ for $S\!=\! 3$ sensors.
%%  and mean number of false alarms $\mu^{(s)} \rmv\!=\! 2$. 
For all methods, as expected, the MOSPA error decreases with decreasing $P_\text{d}$. 
Fig.\ \ref{fig:OSPAvsSensors} shows the time-averaged MOSPA error 
%% (averaged over time steps $n \in \{50,\dots,150\}$) 
versus the number of sensors $S$ for 
%% sensor detection probability 
$P_\text{d} \!=\! 0.6$.
%%  and $\mu^{(s)} \!\rmv=\! 2$. 
It can be seen that the MOSPA error of the MS-PHD and MS-CPHD filters increases for $S$ larger than 5;
this is because the chosen maximum numbers of subsets (120) and partitions (720) 
%% \cite{nannuru15,nannuru15journal} 
are too small for that case. 
(We note that choosing larger maximum numbers of subsets and partitions leads to excessive simulation times.) 
Finally, Fig.\ \ref{fig:OSPAvsClutter} shows the time-averaged MOSPA error versus the mean number of false alarms 
$\mu^{(s)}$ for $S\!=\! 3$ 
%% sensors 
and $P_\text{d} \!=\! 0.6$.  As expected, the MOSPA error of all methods increases with growing $\mu^{(s)}\rmv$. 
In addition, Figs.\ \ref{fig:OSPAvsPD}--\ref{fig:OSPAvsClutter} again show that the proposed method outperforms the other methods.
%% confirm the performance advantage of the proposed BP method over the considered reference
%% state-of-the-art 
%% methods. 
We note that the poor performance of the IC-MB filter is due to the 
%% fact that the 
approximation used by that filter, which is accurate only for a high $P_\text{d}$ and a very low $\mu^{(s)}$
%% number of false alarms 
\cite{vo09}. 
%% For the MOSPA curves in Fig. \ref{fig:OSPAvsPD} and Fig. \ref{fig:OSPAvsSensors}, we performed 400 Monte Carlo run for every entry on the x-axis.

\vspace{-1mm}
 
\subsection{Scalability}
\label{sec:sim-scal}

\begin{figure}
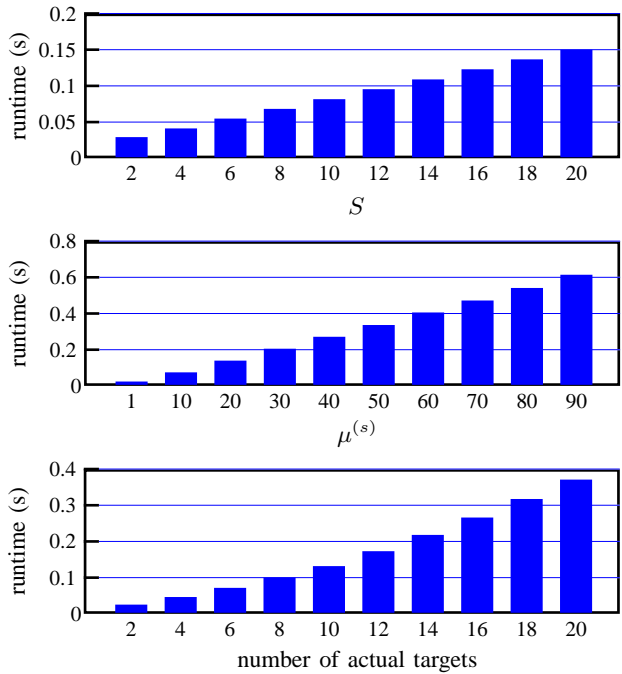

\centering
\footnotesize
\renewcommand{\betweenticks}{0.05}
\renewcommand{\captionoffset}{1.25cm}
\renewcommand{\lefttextoffset}{.5cm}
\renewcommand{\shownumbers}{0}
\bardiagrambegin{13.3}{.2}{0.7cm}{0.9}{1.3}{0.5cm}{9.6cm}
\drawlevellines
\renewcommand{\barlabelangle}{0}
\baritem{2}{0.0307}{blue}
\baritem{4}{0.0427}{blue}
\baritem{6}{0.0563}{blue}
\baritem{8}{0.0697}{blue}
\baritem{10}{0.0831}{blue}
\baritem{12}{0.0969}{blue}
\baritem{14}{0.1104}{blue}
\baritem{16}{0.1244}{blue}
\baritem{18}{0.1381}{blue}
\baritem{20}{0.1515}{blue}
\bardiagramend{{\small \hspace*{-2.5mm}$S$}}{{\small runtime (s)}}
\vspace{4mm}
\renewcommand{\betweenticks}{0.2}
\renewcommand{\captionoffset}{1.25cm}
\renewcommand{\lefttextoffset}{.5cm}
\renewcommand{\shownumbers}{0}
\bardiagrambegin{13.3}{.8}{0.7cm}{0.9}{1.3}{0.5cm}{2.40cm}
\drawlevellines
\renewcommand{\barlabelangle}{0}
\baritem{1}{0.0300}{blue}
\baritem{10}{0.0807}{blue}
\baritem{20}{0.1455}{blue}
\baritem{30}{0.2113}{blue}
\baritem{40}{0.2773}{blue}
\baritem{50}{0.3423}{blue}
\baritem{60}{0.4116}{blue}
\baritem{70}{0.4777}{blue}
\baritem{80}{0.5470}{blue}
\baritem{90}{0.6200}{blue}
\bardiagramend{{\small \hspace*{-2.5mm}$\mu^{(s)}$}}{{\small runtime (s)}}
\vspace{4mm}
\renewcommand{\betweenticks}{0.1}
\renewcommand{\captionoffset}{1.25cm}
\renewcommand{\lefttextoffset}{.5cm}
\renewcommand{\shownumbers}{0}
\bardiagrambegin{13.3}{.4}{0.7cm}{0.9}{1.3}{0.5cm}{4.8cm}
\drawlevellines
\renewcommand{\barlabelangle}{0}
\baritem{2}{0.0284}{blue}
\baritem{4}{0.0494}{blue}
\baritem{6}{0.0747}{blue}
\baritem{8}{0.1035}{blue}
\baritem{10}{0.1348}{blue}
\baritem{12}{0.1760}{blue}
\baritem{14}{0.2210}{blue}
\baritem{16}{0.2690}{blue}
\baritem{18}{0.3202}{blue}
\baritem{20}{0.3738}{blue}
\bardiagramend{{\small \hspace*{-2.5mm}number of actual targets}}{{\small runtime (s)}}
\hspace{3mm}
\vspace{.5mm}
\caption{Average runtime per time step of the proposed method. Top: versus $S$ for five actual
%% intersecting 
targets and $\mu^{(s)} \!\rmv=\! 2$; center:
%% (middle) 
versus $\mu^{(s)}$ for $S \!=\! 3$ and five actual targets; bottom: versus the number of actual targets for $S \!=\! 3$, $\mu^{(s)} \!\rmv=\rmv 2$, and the number of PTs set to $K \!=\rmv 5,7,\ldots,23$.}
\label{fig:numSensorsVsRuntime}
%% \vspace{2mm}
\end{figure}

Finally, we investigate how the runtime of our method scales in
%% with 
the number of sensors $S$ and the number of actual targets, and how it 
depends on the mean number of false alarms $\mu^{(s)}\rmv$. 
Fig.\ \ref{fig:numSensorsVsRuntime} shows the average runtime per time step $n$ versus $S$, $\mu^{(s)}\rmv$, and the number of actual targets
for a MATLAB implementation 
%%  of our method 
on a single core of an Intel Xeon E5-2640 v3 CPU.
The runtime was averaged over 150 time steps and 400 simulation runs. The probability of detection was set to $P_\text{d} \rmv=\rmv 0.6$. 
These results confirm the linear scaling of the runtime in $S$ and in
the mean number of measurements per sensor (which grows linearly with
%% is proportional to 
$\mu^{(s)}$). The scaling in the number of actual targets is 
seen to be roughly quadratic. Further investigation showed that the scaling in the number of actual targets is linear if the number of PTs is held fixed, 
and similarly, the scaling in the number of PTs is linear if the number of actual targets is held fixed. The low absolute complexity of the proposed method is evidenced by the fact that for 20 actual targets, $S \rmv=\rmv 3$ sensors, and $\mu^{(s)} \!=\rmv 2$, the computations per time step $n$ require less than 0.4{\ist}s. 

%% \newpage %%%%%%%

%%%%%%%%%%%%%%%%%%%%%%%%%%%%%%%%%%%%%%%%%%%%%%%%
\section{Conclusion}
%%%%%%%%%%%%%%%%%%%%%%%%%%%%%%%%%%%%%%%%%%%%%%%%%%%%%%%%%%%

We 
%% have 
developed and demonstrated the application of the belief propagation (BP) scheme to the problem of
%% proposed a belief propagation method for 
tracking an unknown number of targets using multiple sensors. The proposed BP-based multitarget tracking method exhibits low complexity and 
excellent scaling properties with respect to all relevant systems parameters. This is achieved through the use of ``augmented target states'' including binary target indicators and
the establishment
%% formulation 
of an appropriate statistical model involving a redundant formulation of data association uncertainty \cite{williams14}. 
%% In particular, the 
The complexity of our method scales only quadratically in the number of targets, linearly in the number of sensors, and linearly in the number of measurements per sensors. 
Simulation results in a challenging scenario with intersecting targets
%% trajectories 
showed that the proposed method can outperform previously proposed methods, including methods with a less favorable scaling behavior. 
In particular, we observed significant improvements in OSPA performance 
%% gains over 
relative to various multisensor extensions of the PHD, CPHD, and multi-Bernoulli filters.

Promising 
%% Possible 
directions for future research include extensions of the proposed BP method that adapt to time-varying environmental conditions, 
e.g., to a time-varying  probability of detection \cite{vo13}, and distributed variants 
%% of the proposed method would be desirable 
for use in 
%% scenarios where the sensors form a 
decentralized wireless sensor networks with communication constraints \cite{sohraby07}. 
%% Finally, a 
A direction of theoretical interest would be
%% from a theoretical perspective, 
a FISST-based derivation of multisensor-multitarget tracking algorithms using BP for data association.
%%  would be interesting.

\bibliographystyle{IEEEtran}
\bibliography{references}

\begin{thebibliography}{10}
\providecommand{\url}[1]{#1}
\csname url@samestyle\endcsname
\providecommand{\newblock}{\relax}
\providecommand{\bibinfo}[2]{#2}
\providecommand{\BIBentrySTDinterwordspacing}{\spaceskip=0pt\relax}
\providecommand{\BIBentryALTinterwordstretchfactor}{4}
\providecommand{\BIBentryALTinterwordspacing}{\spaceskip=\fontdimen2\font plus
\BIBentryALTinterwordstretchfactor\fontdimen3\font minus
  \fontdimen4\font\relax}
\providecommand{\BIBforeignlanguage}[2]{{%
\expandafter\ifx\csname l@#1\endcsname\relax
\typeout{** WARNING: IEEEtran.bst: No hyphenation pattern has been}%
\typeout{** loaded for the language `#1'. Using the pattern for}%
\typeout{** the default language instead.}%
\else
\language=\csname l@#1\endcsname
\fi
#2}}
\providecommand{\BIBdecl}{\relax}
\BIBdecl

\bibitem{koch14}
W.~Koch, \emph{{Tracking and Sensor Data Fusion: Methodological Framework and
  Selected Applications}}.\hskip 1em plus 0.5em minus 0.4em\relax Berlin,
  Germany: Springer, 2014.

\bibitem{wiley15}
B.-N. Vo, M.~Mallick, Y.~Bar-Shalom, S.~Coraluppi, R.~Osborne, R.~Mahler, and
  B.-T. Vo, ``Multitarget tracking,'' in \emph{Wiley Encyclopedia of Electrical
  and Electronics Engineering}, M.~Peterca, Ed.\hskip 1em plus 0.5em minus
  0.4em\relax Hoboken, NJ, USA: Wiley, 2015.

\bibitem{barShalom95}
Y.~Bar-Shalom and X.-R. Li, \emph{{Multitarget-Multisensor Tracking: Principles
  and Techniques}}.\hskip 1em plus 0.5em minus 0.4em\relax Storrs, CT, USA:
  Yaakov Bar-Shalom, 1995.

\bibitem{mahler2007statistical}
R.~Mahler, \emph{Statistical Multisource-Multitarget Information Fusion}.\hskip
  1em plus 0.5em minus 0.4em\relax Norwood, MA, USA: Artech House, 2007.

\bibitem{reid79}
D.~B. Reid, ``An algorithm for tracking multiple targets,'' \emph{IEEE Trans.
  Autom. Control}, vol.~24, no.~6, pp. 843--854, Dec. 1979.

\bibitem{pao95}
L.~Y. Pao and C.~W. Frei, ``A comparison of parallel and sequential
  implementations of a multisensor multitarget tracking algorithm,'' in
  \emph{Proc. ACC-95}, vol.~3, Seattle, WA, USA, Jun. 1995, pp. 1683--1687.

\bibitem{deb97}
S.~Deb, M.~Yeddanapudi, K.~Pattipati, and Y.~Bar-Shalom, ``{A generalized S-D
  assignment algorithm for multisensor-multitarget state estimation},''
  \emph{IEEE Trans. Aerosp. Electron. Syst.}, vol.~33, no.~2, pp. 523--538,
  Apr. 1997.

\bibitem{vermaak05}
J.~Vermaak, S.~J. Godsill, and P.~Perez, ``{Monte Carlo filtering for multi
  target tracking and data association},'' \emph{IEEE Trans. Aerosp. Electron.
  Syst.}, vol.~41, no.~1, pp. 309--332, Jan. 2005.

\bibitem{musicki04}
D.~Musicki and R.~Evans, ``{Joint integrated probabilistic data association:
  JIPDA},'' \emph{IEEE Trans. Aerosp. Electron. Syst.}, vol.~40, no.~3, pp.
  1093--1099, Jul. 2004.

\bibitem{musicki09}
D.~Musicki and R.~J. Evans, ``{Multiscan multitarget tracking in clutter with
  integrated track splitting filter},'' \emph{IEEE Trans. Aerosp. Electron.
  Syst.}, vol.~45, no.~4, pp. 1432--1447, Oct. 2009.

\bibitem{horridge09}
P.~Horridge and S.~Maskell, ``Searching for, initiating and tracking multiple
  targets using existence probabilities,'' in \emph{Proc. FUSION-09}, Seattle,
  WA, USA, Jul. 2009, pp. 611--617.

\bibitem{mahler03}
R.~Mahler, ``{Multitarget Bayes filtering via first-order multitarget
  moments},'' \emph{IEEE Trans. Aerosp. Electron. Syst.}, vol.~39, no.~4, pp.
  1152--1178, Oct. 2003.

\bibitem{vo05}
B.-N. Vo, S.~Singh, and A.~Doucet, ``{Sequential Monte Carlo methods for
  multitarget filtering with random finite sets},'' \emph{IEEE Trans. Aerosp.
  Electron. Syst.}, vol.~41, no.~4, pp. 1224--1245, Oct. 2005.

\bibitem{mahler07}
R.~Mahler, ``{PHD filters of higher order in target number},'' \emph{IEEE
  Trans. Aerosp. Electron. Syst.}, vol.~43, no.~4, pp. 1523--1543, Oct. 2007.

\bibitem{vo07}
B.-T. Vo, B.-N. Vo, and A.~Cantoni, ``Analytic implementations of the
  cardinalized probability hypothesis density filter,'' \emph{IEEE Trans.
  Signal Process.}, vol.~55, no.~7, pp. 3553--3567, Jul. 2007.

\bibitem{ristic13}
B.~Ristic, B.-T. Vo, B.-N. Vo, and A.~Farina, ``{A tutorial on Bernoulli
  filters: Theory, implementation and applications},'' \emph{IEEE Trans. Signal
  Process.}, vol.~61, no.~13, pp. 3406--3430, Jul. 2013.

\bibitem{vo09}
B.-T. Vo, B.-N. Vo, and A.~Cantoni, ``{The cardinality balanced multi-target
  multi-Bernoulli filter and its implementations},'' \emph{IEEE Trans. Signal
  Process.}, vol.~57, no.~2, pp. 409--423, Feb. 2009.

\bibitem{reuter14}
S.~Reuter, B.-T. Vo, B.-N. Vo, and K.~Dietmayer, ``{The labeled multi-Bernoulli
  filter},'' \emph{IEEE Trans. Signal Process.}, vol.~62, no.~12, pp.
  3246--3260, Jun. 2014.

\bibitem{vo14}
B.-N. Vo, B.-T. Vo, and D.~Phung, ``{Labeled random finite sets and the Bayes
  multi-target tracking filter},'' \emph{IEEE Trans. Signal Process.}, vol.~62,
  no.~24, pp. 6554--6567, Dec. 2014.

\bibitem{williams2015marg}
J.~L. Williams, ``{Marginal multi-Bernoulli filters: RFS derivation of MHT,
  JIPDA and association-based MeMBer},'' \emph{IEEE Trans. Aerosp. Electron.
  Syst.}, vol.~51, no.~3, pp. 1664--1687, Jul. 2015.

\bibitem{braca13}
P.~Braca, S.~Marano, V.~Matta, and P.~Willett, ``{Asymptotic efficiency of the
  PHD in multitarget/multisensor estimation},'' \emph{IEEE J. Sel. Topics
  Signal Process.}, vol.~7, no.~3, pp. 553--564, Jun. 2013.

\bibitem{mahler09a}
R.~Mahler, ``{The multisensor PHD filter: I. General solution via multitarget
  calculus},'' in \emph{Proc. SPIE-09}, Orlando, FL, USA, Apr. 2009.

\bibitem{nannuru15}
S.~Nannuru, M.~Coates, M.~Rabbat, and S.~Blouin, ``{General solution and
  approximate implementation of the multisensor multitarget CPHD filter},'' in
  \emph{Proc. IEEE ICASSP-15}, Brisbane, Australia, Apr. 2015, pp. 4055--4059.

\bibitem{nannuru15journal}
------, ``{Multisensor CPHD filter},'' 2016, available online:
  http://arxiv.org/abs/1504.06342.

\bibitem{delande11}
E.~Delande, E.~Duflos, P.~Vanheeghe, and D.~Heurguier, ``{Multi-sensor PHD:
  Construction and implementation by space partitioning},'' in \emph{Proc. IEEE
  ICASSP-11}, Prague, Czech Republic, May 2011, pp. 3632--3635.

\bibitem{nagappa11}
S.~Nagappa and D.~Clark, ``{On the ordering of the sensors in the
  iterated-corrector probability hypothesis density (PHD) filter},'' in
  \emph{Proc. SPIE-11}, Orlando, FL, USA, Apr. 2011, pp. 26--28.

\bibitem{mahler10a}
R.~Mahler, ``{Approximate multisensor CPHD and PHD filters},'' in \emph{Proc.
  FUSION-10}, Edinburgh, UK, Jul. 2010, pp. 26--29.

\bibitem{kschischang01}
F.~R. Kschischang, B.~J. Frey, and H.-A. Loeliger, ``{Factor graphs and the
  sum-product algorithm},'' \emph{IEEE Trans. Inf. Theory}, vol.~47, no.~2, pp.
  498--519, Feb. 2001.

\bibitem{wymeersch07}
H.~Wymeersch, \emph{{Iterative Receiver Design}}.\hskip 1em plus 0.5em minus
  0.4em\relax New York, NY, USA: Cambridge University Press, 2007.

\bibitem{wainwright08}
M.~J. Wainwright and M.~I. Jordan, ``Graphical models, exponential families,
  and variational inference,'' \emph{Found. Trends Mach. Learn.}, vol.~1, Jan.
  2008.

\bibitem{loeliger}
H.-A. Loeliger, ``{An introduction to factor graphs},'' \emph{IEEE Signal
  Process. Mag.}, vol.~21, no.~1, pp. 28--41, Jan. 2004.

\bibitem{chertkov10}
M.~Chertkov, L.~Kroc, F.~Krzakala, M.~Vergassola, and L.~Zdeborov\'{a},
  ``{Inference in particle tracking experiments by passing messages between
  images},'' \emph{{PNAS}}, vol. 107, no.~17, pp. 7663–--7668, Apr. 2010.

\bibitem{williams14}
J.~L. Williams and R.~Lau, ``Approximate evaluation of marginal association
  probabilities with belief propagation,'' \emph{IEEE Trans. Aerosp. Electron.
  Syst.}, vol.~50, no.~4, pp. 2942--2959, Oct. 2014.

\bibitem{chen09}
Z.~Chen, L.~Chen, M.~Cetin, and A.~S. Willsky, ``An efficient message passing
  algorithm for multi-target tracking.'' in \emph{Proc. FUSION-09}, Seattle,
  WA, USA, Jul. 2009, pp. 826--833.

\bibitem{meyer15scalable}
F.~Meyer, P.~Braca, P.~Willett, and F.~Hlawatsch, ``{Scalable multitarget
  tracking using multiple sensors: A belief propagation approach},'' in
  \emph{Proc. FUSION-15}, Washington D.C., USA, Jul. 2015, pp. 1778--1785.

\bibitem{horridge06}
P.~Horridge and S.~Maskell, ``{Real-time tracking of hundreds of targets with
  efficient exact JPDAF implementation},'' in \emph{Proc. FUSION-06}, Florence,
  Italy, Jul. 2006, pp. 1--8.

\bibitem{chen06}
L.~Chen, M.~J. Wainwright, M.~Cetin, and A.~S. Willsky, ``{Data association
  based on optimization in graphical models with application to sensor
  networks},'' \emph{Math. Comp. Model.}, vol.~43, no. 9--10, pp. 1114--1135,
  2006.

\bibitem{meyer16scalable}
F.~Meyer, P.~Braca, P.~Willett, and F.~Hlawatsch, ``{Tracking an unknown number
  of targets using multiple sensors: A belief propagation method},'' in
  \emph{Proc. FUSION-16}, Heidelberg, Germany, Jul. 2016.

\bibitem{poor94}
H.~V. Poor, \emph{{An Introduction to Signal Detection and Estimation}}.\hskip
  1em plus 0.5em minus 0.4em\relax New York, NY: Springer, 1994.

\bibitem{kay1993}
S.~M. Kay, \emph{\BIBforeignlanguage{english}{{Fundamentals of Statistical
  Signal Processing: Estimation Theory}}}.\hskip 1em plus 0.5em minus
  0.4em\relax Upper Saddle River, NJ, USA: Prentice-Hall, 1993.

\bibitem{wymeersch09}
H.~Wymeersch, J.~Lien, and M.~Z. Win, ``{Cooperative localization in wireless
  networks},'' \emph{Proc. IEEE}, vol.~97, no.~2, pp. 427--450, Feb. 2009.

\bibitem{vontobel13}
P.~O. Vontobel, ``{The Bethe permanent of a nonnegative matrix},'' \emph{IEEE
  Trans. Inf. Theory}, vol.~59, no.~3, pp. 1866--1901, Mar. 2013.

\bibitem{arulampalam01}
M.~S. Arulampalam, S.~Maskell, N.~Gordon, and T.~Clapp, ``{A tutorial on
  particle filters for online nonlinear/non-Gaussian Bayesian tracking},''
  \emph{IEEE Trans. Signal Process.}, vol.~50, no.~2, pp. 174--188, Feb. 2002.

\bibitem{doucet01}
A.~Doucet, N.~de~Freitas, and N.~Gordon, \emph{{Sequential Monte Carlo Methods
  in Practice}}.\hskip 1em plus 0.5em minus 0.4em\relax New York, NY, USA:
  Springer, 2001.

\bibitem{ristic12}
B.~Ristic and S.~Arulampalam, ``{Bernoulli particle filter with observer
  control for bearings-only tracking in clutter},'' \emph{IEEE Trans. Aerosp.
  Electron. Syst.}, vol.~48, no.~3, pp. 2405--2415, Jul. 2012.

\bibitem{kropfreiter16}
T.~Kropfreiter, F.~Meyer, and F.~Hlawatsch, ``{Sequential Monte Carlo
  implementation of the track-oriented marginal multi-Bernoulli/Poisson
  filter},'' in \emph{Proc. FUSION-16}, Heidelberg, Germany, Jul. 2016.

\bibitem{kotecha03}
J.~H. Kotecha and P.~M. Djuric, ``{Gaussian particle filtering},'' \emph{IEEE
  Trans. Signal Process.}, vol.~51, no.~10, pp. 2592--2601, Oct 2003.

\bibitem{mahler09b}
R.~Mahler, ``{The multisensor PHD filter: II. Erroneous solution via Poisson
  magic},'' in \emph{Proc. SPIE-09}, Orlando, FL, USA, Apr. 2009.

\bibitem{schuhmacher08}
D.~Schuhmacher, B.-T. Vo, and B.-N. Vo, ``A consistent metric for performance
  evaluation of multi-object filters,'' \emph{IEEE Trans. Signal Process.},
  vol.~56, no.~8, pp. 3447--3457, Aug. 2008.

\bibitem{gan07}
G.~Gan, C.~Ma, and J.~Wu, \emph{{Data Clustering: Theory, Algorithms, and
  Applications}}.\hskip 1em plus 0.5em minus 0.4em\relax Philadelphia, PA, USA:
  SIAM, 2007.

\bibitem{vo13}
B.-T. Vo, B.-N. Vo, R.~Hoseinnezhad, and R.~Mahler, ``{Robust multi-Bernoulli
  filtering},'' \emph{IEEE J. Sel. Topics Signal Process.}, vol.~7, no.~3, pp.
  399--409, Jun. 2013.

\bibitem{sohraby07}
K.~Sohraby, D.~Minoli, and T.~Znati, \emph{{Wireless Sensor Networks:
  Technology, Protocols, and Applications}}.\hskip 1em plus 0.5em minus
  0.4em\relax Hoboken, NJ, USA: Wiley, 2007.

\end{thebibliography}

\end{document}